\documentclass[twocolumn,11pt]{aastex63}
\usepackage{url}
\usepackage{gensymb}
\usepackage{textcomp}
\usepackage{xspace}
\usepackage{natbib,amsmath} \usepackage{graphicx}
\usepackage{hyperref} \usepackage{float} \usepackage{subfigure}
\usepackage{tabularx}
\providecommand{\e}[1]{\ensuremath{\times 10^{#1}}}
\usepackage{bm}

\newcommand{\ang}{\AA\xspace}
\newcommand{\dust}{\mathrm{dust}}

\newcommand*\chem[1]{\ensuremath{\mathrm{#1}}}

\newcommand{\cm}{\mathrm{cm}}
\newcommand{\g}{\mathrm{g}}
\newcommand{\K}{\mathrm{K}}  
\newcommand{\eV}{\mathrm{eV}}
\newcommand{\keV}{\mathrm{keV}}
\newcommand{\s}{\mathrm{s}}

\newcommand{\p}{\mathrm{p}}

\usepackage{array}

\begin{document}
\title{Detection of Ongoing Mass Loss from HD 63433c, a Young Mini Neptune}

\correspondingauthor{Michael Zhang}
\email{mzzhang2014@gmail.com}

\author[0000-0002-0659-1783]{Michael Zhang}
\affiliation{Department of Astronomy, California Institute of Technology, Pasadena, CA 91125, USA}

\author[0000-0002-5375-4725]{Heather A. Knutson}
\affiliation{Division of Geological and Planetary Sciences, California Institute of Technology}

\author[0000-0002-6540-7042]{Lile Wang}
\affiliation{Center for Computational Astrophysics, Flatiron Institute, New York, NY 10010, USA}

\author[0000-0002-8958-0683]{Fei Dai}
\affiliation{Division of Geological and Planetary Sciences, California Institute of Technology}

\author[0000-0002-2248-3838]{Leonardo A. dos Santos}
\affiliation{Space Telescope Science Institute, 3700 San Martin Drive, Baltimore, MD 21218, USA}
\affiliation{Observatoire astronomique de l’Université de Genève, Chemin Pegasi 51, 1290 Versoix, Switzerland}

\author[0000-0003-4426-9530]{Luca Fossati}
\affiliation{Space Research Institute, Austrian Academy of Sciences, Schmiedlstrasse 6, A-8042, Graz, Austria}

\author[0000-0003-4155-8513]{Gregory W. Henry}
\affiliation{Center of Excellence in Information Systems, Tennessee State University, Nashville, TN  37209, USA}

\author[0000-0001-9704-5405]{David Ehrenreich}
\affiliation{Observatoire astronomique de l’Université de Genève, Chemin Pegasi 51, 1290 Versoix, Switzerland}

\author[0000-0002-4644-8818]{Yann Alibert}
\affiliation{Universitat Bern}

\author[0000-0003-3477-2466]{Sergio Hoyer}
\affiliation{Aix Marseille Univ, CNRS, CNES, LAM, Marseille, France}

\author[0000-0001-8749-1962]{Thomas G. Wilson}
\affiliation{SUPA, School of Physics and Astronomy, University of St. Andrews, North Haugh, Fife KY16 9SS, UK}

\author{Andrea Bonfanti}
\affiliation{Space Research Institute, Austrian Academy of Sciences, Schmiedlstrasse 6, A-8042, Graz, Austria}

\begin{abstract}
We detect Lyman $\alpha$ absorption from the escaping atmosphere of HD 63433c, a $R=2.67 R_\Earth$, $P=20.5$ d mini Neptune orbiting a young (440 Myr) solar analogue in the Ursa Major Moving Group.  Using HST/STIS, we measure a transit depth of $11.1 \pm 1.5$\% in the blue wing and $8 \pm 3$\% in the red.  This signal is unlikely to be due to stellar variability, but should be confirmed by an upcoming second transit observation with HST.  We do not detect Lyman $\alpha$ absorption from the inner planet, a smaller $R=2.15 R_\Earth$ mini Neptune on a 7.1 d orbit.  We use Keck/NIRSPEC to place an upper limit of 0.5\% on helium absorption for both planets.  We measure the host star's X-ray spectrum and MUV flux with XMM-Newton, and model the outflow from both planets using a 3D hydrodynamic code.  This model provides a reasonable match to the light curve in the blue wing of the Lyman $\alpha$ line and the helium non-detection for planet c, although it does not explain the tentative red wing absorption or reproduce the excess absorption spectrum in detail.  Its predictions of strong Lyman $\alpha$ and helium absorption from b are ruled out by the observations.  This model predicts a much shorter mass loss timescale for planet b, suggesting that b and c are fundamentally different: while the latter still retains its hydrogen/helium envelope, the former has likely lost its primordial atmosphere.
\end{abstract}


\section{Introduction}
\label{sec:introduction}
Mass loss shapes exoplanet demographics and atmospheric properties.  The observed radius distribution of close-in sub-Neptune-sized planets is bimodal, with peaks at $<$ 1.5 R$_\Earth$ and 2-3 R$_\Earth$ \citep{fulton_2017,fulton_2018}.  This bimodality can be explained by scenarios in which the observed population of sub-Neptune-sized planets formed with a few M$_\texttt{Earth}$ rocky cores and hydrogen-rich atmospheres, which were then stripped away from the most highly irradiated planets (e.g., \citealt{lopez_2013, owen_2017, lehmer_2017, mills_2017}).  However, it is possible that the gap is caused by mass loss driven by the forming protoplanet's own cooling luminosity \citep{ginzburg_2018,gupta_2019}.  It has also been proposed that the radius valley is primordial \citep{lee_2021}.

An escaping atmosphere can be detected in absorption when the planet transits in front of its host star.  Because of the abundance of hydrogen and the strength of the Ly$\alpha$ line, Ly$\alpha$ exospheres can absorb a very large fraction of starlight during transit.  The Neptune-sized GJ 436b, for example, has a transit depth of 56\% in the Ly$\alpha$ blue wing and 47\% in the red wing \citep{lavie_2017}.  The second most abundant element of escaping primordial atmospheres is helium.  In 2018, \cite{spake_2018} detected helium absorption from a transiting exoplanet for the first time.  The He I 1083 nm line is observable from the ground and has a much higher photon flux compared to Ly$\alpha$. It also has its own challenges: the signal size is much smaller, and only active early M to late G stars have the right high-energy spectrum to maintain a suitably high triplet ground state population \citep{oklopcic_2018}.  These two probes provide complementary information.  Because ISM absorption wipes out the Ly$\alpha$ core, Ly$\alpha$ probes high-velocity hydrogen in the tenuous outer reaches of the escaping atmosphere.  The metastable helium line is not absorbed by the ISM and probes low-velocity helium in the denser part of the exosphere, closer to the planet surface ($\sim$3 planetary radii vs. $\sim$12 planetary radii).

The young mini Neptune regime is the most critical for understanding the processes behind the radius gap, yet mass loss has never been securely detected from planets of this size in either Ly$\alpha$ or helium.  The smallest planet with a secure detection in either wavelength is the Neptune-mass GJ 3470b, with a radius of 3.9 $R_\Earth$ and a mass of 13 $M_\Earth$\citep{bourrier_2018b}.  However, it is not young, with a rotation period of 20 days and an age of 1--4 Gyr \citep{biddle_2014}.  The other planets with definite Ly$\alpha$ detecions--GJ 436b \citep{lavie_2017}, HD 189733b\citep{bourrier_2013}, HD 209458b \citep{vidal-madjar_2008}--are even bigger and older.  Of the planets detected in helium, none are mini Neptunes or smaller, and none are indisputably younger than 1 Gyr.  WASP-107b \citep{mocnik_2017} has a gyrochronological age of $0.6 \pm 0.2$ Gyr but an evolutionary track age estimate of $8.3 \pm 4.3$ Gyr, illustrating the difficulty of measuring ages for isolated stars.

The scarcity of successful measurements is due to the many conditions necessary for Ly$\alpha$ absorption or helium absorption to be detectable.  For both wavelengths, we need a transiting exoplanet with a hydrogen-rich atmosphere on a tight orbit around a star of at least moderate activity.  Interstellar Ly$\alpha$ absorption saturates for even the closest stars, making it hard to observe planetary absorption beyond 50 pc and almost impossible beyond 100 pc.  Triplet helium absorption requires a high population of triplet helium, which in turn requires a star that has a high extreme UV to mid UV ratio, which is optimally achieved for K type stars, but not impossible around active G stars. Very few currently known transiting planets fit these criteria, and few of those are young mini Neptunes.

The G5 star HD 63433 (TOI 1726) is a young ($414 \pm 23$ Myr) and nearby (22 pc) solar analogue (M=1 $M_{\odot}$), a member of the Ursa Major moving group \citep{mann_2020}.  In keeping with its young age and 6.4 d rotation period, it has an exceptionally high X-ray luminosity.  The Second ROSAT All-sky Survey measured its 0.1--2.4 KeV X-ray flux to be $F_{\mathrm{X}} = 1.4-1.8 \times 10^{-12}$ erg s$^{-1}$ cm$^{-2}$.  The XMM-Newton slew survey \citep{saxton_2008} measured its 0.2--12 KeV flux as $4.4 \pm 1.4$ and $1.6 \pm 0.8$  erg s$^{-1}$ cm$^{-2}$ on two separate visits.  Taking 1.5\e{-12} to be representative, the corresponding stellar X-ray luminosity is $10^{29}$ erg s$^{-1}$, 40 times higher than the average solar X-ray luminosity \citep{judge_2003}.  In addition, the star's negative radial velocity ($-16$ km/s), together with the positive radial velocity of the local interstellar cloud in that direction (22 km/s), gives an unusually clear view of the blue Ly$\alpha$ wing and a glimpse of the core.  Luckily, the core and blue wing are where we expect the most planetary absorption: planetary outflows have a typical speed of $\sim$2 times the sound speed (or $\sim$20 km/s), and the stellar wind pushes the outflow toward the observer.

Inside this intense X-ray environment reside two mini Neptunes, both discovered by TESS \citep{mann_2020}: a 2.15 $R_\Earth$ planet on a 7.1 day orbit, and a 2.67 $R_\Earth$ planet on a 20.5 day orbit.  Although they do not have measured masses, previous radial velocity and transit timing studies have found that even mature planets in this size range typically have low densities consistent with the presence of volatile-rich envelopes (e.g., \citealt{rogers_2015,wolfgang_2015,hadden_2017}).  If these planets do have hydrogen/helium envelopes, their young age and the high X-ray luminosity of their host star point to the likelihood of ongoing mass loss.  HD 63443b, in particular, is at an orbital period where there are more super Earths (1-1.5 $R_\Earth$) than mini Neptunes (2-3 $R_\Earth$), as shown in Figure 6 of \cite{fulton_2018}.  Although it may currently have a gaseous envelope, this envelope is likely to be stripped away, moving it into the super Earth population.

HD 63433 is a uniquely favorable target for mass loss studies.  Its young age, high activity, negative radial velocity, and close-in mini Neptunes provide ideal conditions for probing mass loss in the most critical regime.  The existence of two mini Neptunes in the same system allows us to test hydrodynamical models by comparing their predictions for the two planets to observations: a comparative approach that has hitherto been impossible.  No closer transiting mini Neptune host younger than 1 Gyr is known, let alone one with the other desirable properties to boot.

To study this system, we marshalled a variety of space and ground telescopes to characterize the star's high-energy spectrum and look for absorption from the escaping upper atmospheres in the Ly$\alpha$ line and the helium line.  We describe our observations and data reduction in Section \ref{sec:data_reduction}, our analysis in Section \ref{sec:analysis}, our modelling of the star in Section \ref{sec:understanding_star}, and our modelling of the planetary exospheres in Section \ref{sec:modeling}.  After comparing models to observations and discussing the broader context of our work in Section \ref{sec:discussion}, we conclude in Section \ref{sec:conclusion}.

\section{Observations and Data reduction}
\label{sec:data_reduction}

We characterize the extended atmospheres and corresponding present-day mass loss rates for both planets by measuring the wavelength-dependent transit depth when the planet passes in front of its host star.  We observe transits of both planets in the Ly$\alpha$ line with the Space Telescope Imaging Spectrograph on \emph{Hubble Space Telescope} (\emph{HST}/STIS) \citep{woodgate_1998}, and in the 1083 nm helium triplet with the updated NIRSPEC on Keck \citep{martin_2018}.  We then compare the measured absorption during transit to predictions from mass loss models for each planet.  In order to create these models, we must have a good knowledge of the high energy spectrum of the host star, which drives the outflows in our models.  We use XMM-Newton to characterize the X-ray spectrum of the star, and estimate its extreme UV spectrum using scaling relations based on the reconstructed stellar Ly$\alpha$ emission flux.  We also use archival data from ROSAT, which observed the star in 1990 as part of the ROSAT All-Sky Survey, and optical monitoring data from the T3 0.40m Automatic Photoelectric Telescope (APT) at Fairborn Observatory, to characterize the star's long-term variability and activity cycle.

\subsection{HST/STIS}
With HST/STIS, we obtained two 9-orbit transit observations of the Ly$\alpha$ line with the MAMA detector (program 16319, PI: Michael Zhang).  Because the South Atlantic Anomaly (SAA) prevents more than 5-6 consecutive orbits of observation, we observe as many orbits as we can in the vicinity of the transit, take no data for the 7-9 South Atlantic Anomaly-affected orbits, and observe the remaining 3-4 orbits after the gap.  All orbits except the first contain 2523 s of science exposure time in TIME-TAG mode using the G140M grism with a central wavelength of 1222 \AA{} and a slit width of 52 x 0.1$\arcsec$.  The first orbit in the pre-gap and post-gap segments contain only 1515 s of science exposure time because target acquisition and acquisition peak-up occur during these orbits.  In all orbits, wavelength calibration occurs after the science exposure, during occultation.

On Oct 29/30, 2020 UTC, HST observed a transit of planet c, with 5 orbits near transit and 4 orbits after the gap.  On Jan 28/29, 2021 UTC, it observed a transit of planet b with the same configuration.  On Mar 19, 2021, it attempted to observe six consecutive orbits bracketing a second transit of b, but STIS entered safe mode before the observations were to begin and all data were lost.  On the following day, it successfully observed a 3-orbit baseline.

For our analysis of these data, we rely on \texttt{stistools} 1.3.0, a Python package provided by Space Telescope Science Institute that contains several relevant functions for data reduction.  We start with the tag files, lists of photons that encode the time of arrival and position on the detector.  First, we use \texttt{inttag} to turn the photon lists into raw images by accumulating the photons into 315 second subexposures (303 s for the first orbit).  The first orbit of every visit contains 5 subexposures, while subsequent visits contain 8.  The photon wavelengths are Doppler corrected prior to accumulation to account for HST's orbit around the Earth.

Second, we use \texttt{calstis} to perform standard data reduction tasks,  including subtracting the dark image, flat fielding, rejecting cosmic rays, and wavelength calibration.  Wavelength calibration is performed using the wavelength calibration files taken during occultation, which contain lamp lines but no astrophysical signal.  These files allow \texttt{wavecal}, a component of \texttt{calstis}, to assign a wavelength to every pixel.

The last step is spectral extraction.  \texttt{x1d} attempts to locate the spectrum in the spatial direction but often fails because it excludes the region around the Ly$\alpha$ line to remove geocoronal emission--a process which also removes almost all stellar flux.  Instead, we locate the spectrum ourselves by summing the columns in each row, subtracting a smoothed version of the sums to remove skyglow variations, and fitting a Gaussian to the 30 pixels closest to the peak.  After receiving the spectrum location, \texttt{x1d} sums up the pixel values in extraction windows 1 pixel wide and 11 pixels high, centered on the computed trace location.  To compute the background, it uses two 5 pixel high windows, 40 pixels from the trace on either side.  Unlike the spectral extraction window, the background extraction windows are tilted to account for the tilt of the iso-wavelength contours.  Finally, \texttt{x1d} subtracts the background from the gross flux to get the net flux.

After \texttt{x1d} extracts the spectrum for every subexposure, we interpolate the spectrum onto a common wavelength grid for all subexposures.  The grid has a linear spacing of 0.053 \AA, matching the pixel scale of the detector.

\subsection{Keck/NIRSPEC}
\label{subsec:keck_reduction}
Our NIRSPEC/Keck data (program C261) were taken on Dec 30, 2020 (transit of planet c) and Jan 7, 2021 (transit of planet b).  All observations were in $Y$ band in the high resolution mode.  On Dec 30, the transit was in progress when our observations started.  We collected 3.4 h of in-transit observations and 5 hours of post-transit baseline.  On Jan 7, the transit started toward the end of our observations.  We collected 5.4 h of pre-transit and 2.6 h of in-transit observations.  On both nights, we used the 12 x 0.432$\arcsec$ slit, giving the spectrograph a resolution of 25,000.  The sky was clear, and the seeing (1--1.5$\arcsec$) was poor but typical of this time of year.  In addition, the telescope suffered from wind shake on Dec 30, further broadening the line profile in the spatial direction.  We achieved a typical SNR of 400 per spectral pixel in 60-second exposures, all taken in the ABBA nod pattern to eliminate background.  Because we only used one coadd per exposure, we achieved a high observation efficiency of 77\%.

We calibrated the raw images and extracted 1D spectra for each order using a custom Python pipeline designed for the upgraded NIRSPEC.  The pipeline is described in detail in \cite{zhang_2020}, but we summarize it here.  First, we subtract crosstalk from each raw frame.  Then, we create a master flat, identifying bad pixels in the process.   We use this master flat to compute a calibrated A-B difference image for each A/B pair.  After identifying the spectral trace containing the 1083 nm lines, we use optimal spectral extraction to obtain 1D spectra along with their errors.  We create a template from model telluric lines and a model stellar spectrum, shifted in wavelength to account for the star's average Earth-relative radial velocity during that night.  We then use this template to derive the wavelength solution for each individual spectrum.  

After extracting the 1D spectra, we place the data from each night on a uniform wavelength grid and remove signals not related to the planet.  We do this using SYSREM, which can be thought of as Principal Component Analysis with error bars \citep{mazeh_2007}.  After removing the first principal component, we shift to the planetary frame, divide the data up into in-transit and out-of-transit portions and compare the portions to search for planetary absorption.  Removing more principal components worsens the self-subtraction problem, already severe with a single component (see Section \ref{sec:analysis}).

\subsection{XMM-Newton}
\label{subsec:xmm_data_reduction}
On March 26 2021, XMM-Newton observed the star for 6 ks (XMM prop. ID 088287, PI: Michael Zhang).  XMM-Newton has 3 European Photon Imaging Camera (EPIC) detectors with different technologies (2 MOS, 1 pn), 2 Reflection Grating Spectrometers, and an Optical Monitor, all of which observe the target simultaneously.  We configured the EPIC cameras to observe with the medium filter and small window, giving us 97\% observing efficiency on the two MOS CCDs and 71\% efficiency on the one pn CCD.  These observations measure the star's X-ray spectrum, which plays an important role in driving photoevaporative mass loss.  We configured the Optical Monitor to observe the star in the UVM2 filter ($\lambda=231 \pm 48$ nm) for 2.7 ks and the UVW2 filter ($\lambda=212 \pm 50$ nm) for 2.9 ks.  These observations measure the star's mid ultraviolet flux, which can photoionize metastable helium but not create it, and therefore tend to decrease helium absorption in the metastable 1083 nm line.  Although these observations are not simultaneous with the Ly$\alpha$ and helium mass loss observations, they are within 6 months, while the $P_{\rm cyc}/P_{\rm rot}$ vs. $P_{\rm rot}$ relationship derived by \cite{mascareno_2016} implies a stellar cycle period of 5.5 years (with $\sim40\%$ uncertainty).

To analyze XMM-Newton data, we download the raw Observation Data File (ODF) and use the Science Analysis System (SAS)\footnote{\url{https://www.cosmos.esa.int/web/xmm-newton/sas}} provided by the XMM-Newton team to reduce it.  We run \texttt{xmmextractor}, thereby going from ODF to spectra with default settings and no human intervention.  For the Optical Monitor, SAS produces the light curve of the star in the UVW2 and UVM2 filters, the two mid ultraviolet filters we selected.  For each of the two Reflection Grating Spectrometers (RGSs), SAS produces two spectra (first and second order) and other data products, which we do not use because RGS has only a tenth of the throughput of the EPIC detectors in addition to substantial background; no stellar signal is visible in the data.  For each of the three EPIC detectors, SAS generates the light curve, the background-subtracted spectrum, the Redistribution Matrix File (RMF), and the Ancillary Response File (ARF).  The ARF gives the effective area of the detector as a function of photon energy, while the RMF gives the probability of a photon being detected in each channel as a function of photon energy.  In optical astronomy terminology, the ARF gives the throughput multiplied by aperture area, while the RMF gives the wavelength-dependent line spread profile.

With XMM, as with ROSAT, the RMF is nearly diagonal for high energies, but is far from diagonal at low energies, where most of the stellar flux resides.  In addition, the ARF is highly energy dependent for the two EPIC MOS detectors, although not for the pn-detector.  These factors mean it is impossible to simply plot the measurements and see what the X-ray spectrum looks like.  Rather, it is necessary to have a forward model and fit the parameters to find the best match to the data, taking into account the RMF and ARF.  To do this fitting, we use the interactive tool \texttt{xspec} 12.11.1 \citep{arnaud_1996}.

\subsection{ROSAT}
To analyze the ROSAT All-Sky Survey (RASS) data for HD 63433, we download the data for the relevant sector\footnote{\url{https://heasarc.gsfc.nasa.gov/FTP/rosat/data/pspc/processed_data/900000/rs931219n00/}} and reduce it using HEAsoft 6.28\footnote{\url{https://heasarc.gsfc.nasa.gov/docs/software/lheasoft/}} by following the guide ``ROSAT data analysis using \texttt{xselect} and FTOOls''\footnote{\url{https://heasarc.gsfc.nasa.gov/docs/rosat/ros_xselect_guide/xselect_ftools.html}}.  We define the source region to be a circle centered on the source with a radius of 200 arcseconds.  Following the advice of \cite{belloni_1994}, we define two circular background regions on either side of the source along the scan direction, both 800 arcseconds away and with a radius of 200 arcseconds.  Using \texttt{xselect}, we extract the source spectrum and the background spectrum from the events list.  We download the RMF for the PSPC-C detector\footnote{\url{https://heasarc.gsfc.nasa.gov/docs/rosat/pspc_matrices.html}} and use \texttt{pcarf} (part of the ROSAT subpackage of FTOOLS) to generate the ARF.  The image file has negative EXPOSURE, DETC (deadtime correction), and ONTIME header values to indicate that there is no unique value: the image is pieced together from scanning observations, and the effective exposure time is different at each pixel.  We use the exposure map to determine the correct exposure time (437 s), and set the correct EXPOSURE, DETC, and ONTIME on the source and background files.  Finally, we use \texttt{xspec} to load the source, background, RMF, and ARF, and analyze the data in the same way as the XMM observations.

\section{Analysis of Transit Data}
\label{sec:analysis}
\subsection{New ephemerides}
As part of a CHEOPS Guaranteed Time Observation (GTO) program  CH\_PR100031, one transit of each planet was observed on UT Dec 9/10, 2020 (c) and UT Nov 25/26, 2020 (b).  We combine these data with sector 20 TESS observations taken from December 24, 2019, to January 20, 2020 to refine the ephemeris so that we can predict the transit midpoint to $\sim$1 minute accuracy at the time of our HST and Keck observations.  The new ephemeris is significantly more precise than the old TESS-only estimate, which had an accuracy of $\sim$30 min during the epochs of our HST and Keck observations. For c, we obtain $T_0=2458844.05824 \pm 0.00048$ BJD and $P=20.543888^{+0.000046}_{-0.000045}$ d.  For b, we obtain $T_0 = 2458916.45142^{+0.00030}_{-0.00032}$ BJD and $P = 7.107789 \pm 0.000010$ d.  

\subsection{Ly$\alpha$ absorption during transit}
\label{subsec:ly_alpha_analysis}
\begin{figure*}
  \centering \subfigure {\includegraphics
    [width=0.33\textwidth]{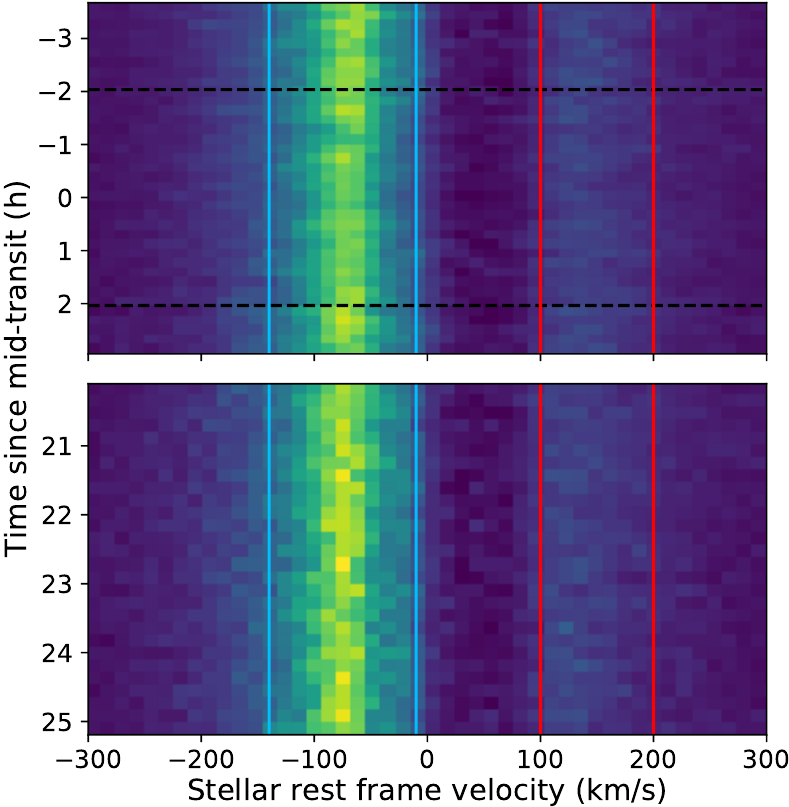}}\subfigure {\includegraphics
    [width=0.33\textwidth]{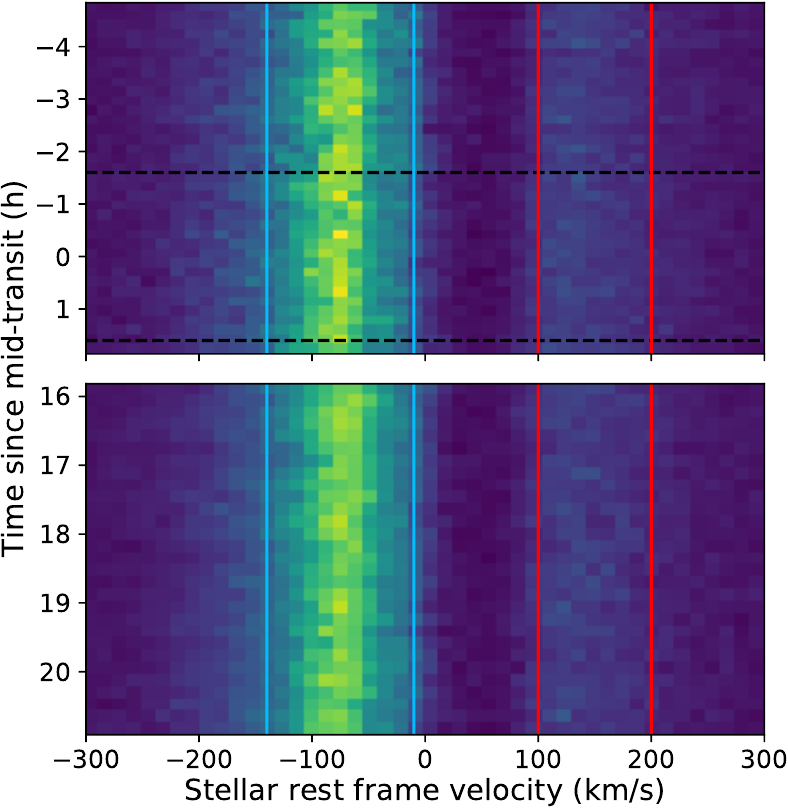}}\subfigure {\includegraphics
    [width=0.33\textwidth]{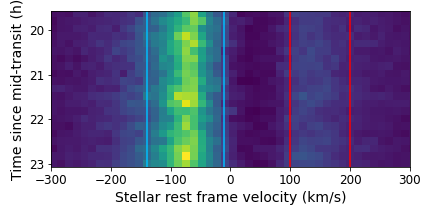}}
    \caption{HST/STIS spectra for planet c (left) and planet b (center and right for the earlier and later observations).  Flux increases from dark blue to yellow.  The white horizontal gap represents the 7--9 orbit hiatus due to the South Atlantic Anomaly (SAA) crossing.  Gaps due to Earth occultation are not shown to avoid visual distraction; the y axis is therefore not strictly accurate because it assumes time increases uniformly from spectrum to spectrum.  The horizontal dashed black lines indicate the beginning and end of the white light transit.  The region between the blue (red) vertical lines is what we define as the blue (red) wing.  They correspond to [$-140$,$-10$] km/s and [100,200] km/s.}
\label{fig:heatmaps}
\end{figure*}

We first examine the UV data to search for signs of Ly$\alpha$ absorption during the transits of planets b and c.  In Figure \ref{fig:heatmaps}, we show the spectral sequence from each HST visit.  For planet c, a clear decrease in the blue wing flux can be seen during the planetary transit.  For planet b, the blue wing does not appear markedly different during transit.  The red wing is more than four times dimmer than the blue wing, making it harder to see any planetary absorption in these 2D plots. 

\begin{figure*}
  \centering \subfigure {\includegraphics
    [width=0.5\textwidth]{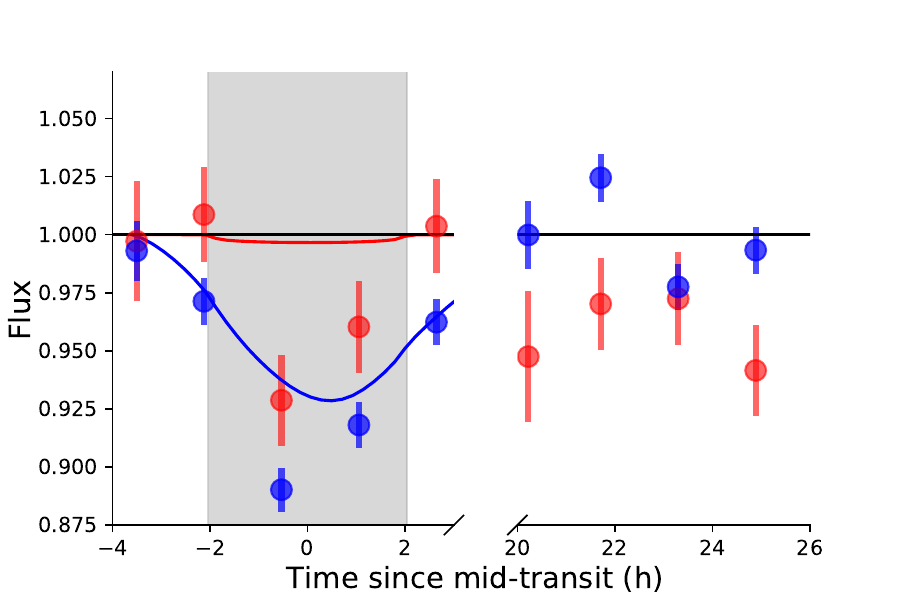}}\subfigure {\includegraphics
    [width=0.5\textwidth]{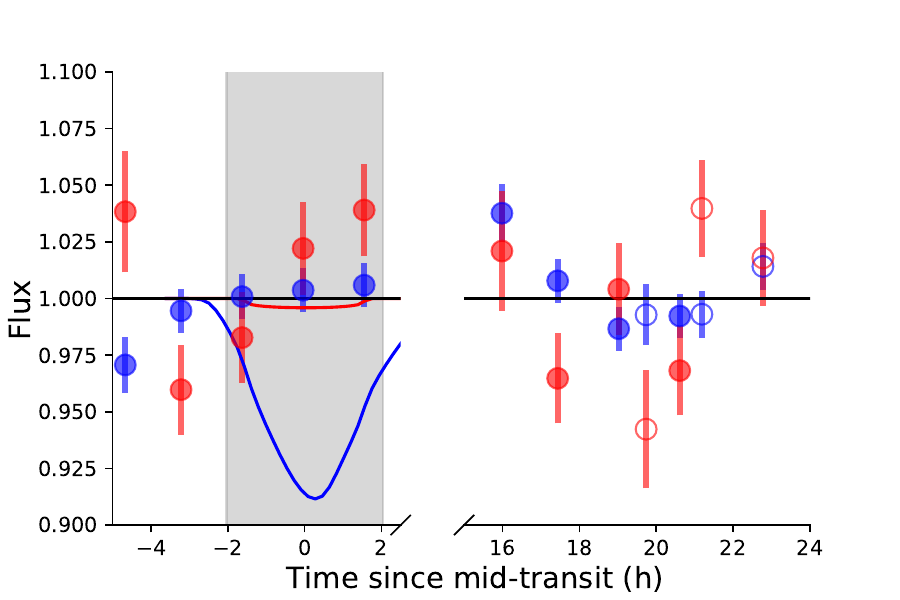}}
    \caption{Ly$\alpha$ light curves for planet c (left) and b (right), in the red and blue wings, compared to the predictions from our 3D hydrodynamic model.  For b, the solid circles represent the first visit, and the open circles represent the failed second visit.  The error bar for each orbit is its photon noise.  The grey region represents the white light transit.}
\label{fig:lyman_alpha_lc}
\end{figure*}

\begin{figure}
    \subfigure{\includegraphics[width=0.5\textwidth]{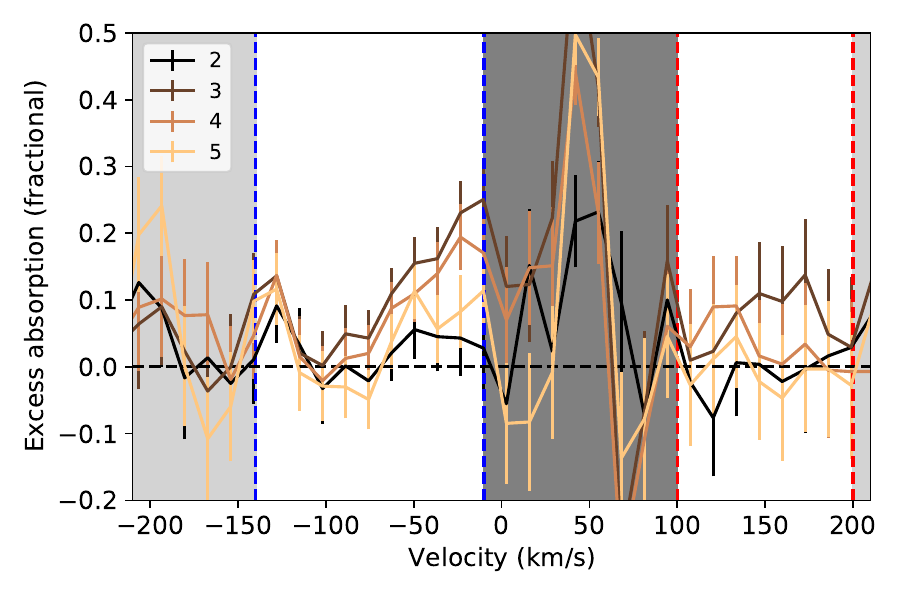}}\quad\subfigure {\includegraphics[width=0.5\textwidth]{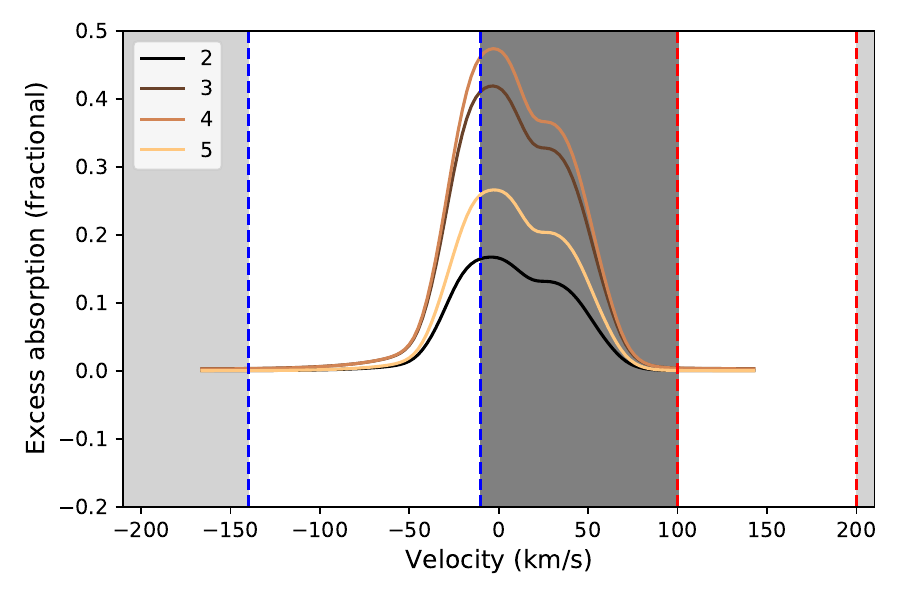}}
    \caption{Per-orbit excess absorption spectrum for planet c, in the observations (top) and the fiducial model (bottom).  The first orbit is used to compute the out-of-transit spectrum.  The white regions are the blue and red wings, the light grey regions are the far wings, and the dark grey region has low flux because of interstellar absorption.  The second and third orbits, in addition to the second half of the second orbit, are within the white light transit.}
    \label{fig:excess_spectrum_c}
\end{figure}

Previous studies have established that HST Ly$\alpha$ observations exhibit modulations in flux within each orbit, which has been attributed to telescope breathing (e.g. \citealt{kimble_1998,ehrenreich_2015}).  According to \cite{kimble_1998}, thermal variations over the course of each spacecraft orbit move the secondary mirror, which changes the focus, which leads to 10-20\% variations in slit loss for the smallest slits.  However, 10-20\% variations in flux are observed for Ly $\alpha$ data taken in the 0.05\arcsec~slit (e.g. \citealt{ehrenreich_2015}), the 0.1\arcsec~slit (e.g. most observations in \citealt{lavie_2017}), and the 0.2\arcsec~slit (e.g. \citealt{munoz_2020}).  This insensitivity to slit size indicates that intra-orbit flux variations are probably not the primary cause of the breathing effect.

We attempt to correct for the instrumental flux variations by decorrelating against a variety of different variables, including time since beginning of orbit, the centroid position of the blue wing, the latitude and longitude of the telescope, and the focus of the telescope as estimated from temperature sensors\footnote{\url{https://www.stsci.edu/hst/instrumentation/focus-and-pointing/focus/hst-focus-model}}.  However, we find that none of these instrumental noise models can reduce the scatter in our light curves.  Although there is clearly a correlation between flux and orbital phase in our observations, the shape of this trend varies from orbit to orbit.  The measured flux increases steeply with time for the first orbit in a visit, but the increase becomes less pronounced in future orbits, until it becomes flat in the fourth or fifth orbits.  As a result, we cannot remove this effect by detrending with a simple function of orbital phase as other papers do (e.g. \citealt{bourrier_2013}).

Since we are unable to effectively remove these intra-orbit flux variations, we instead bin our light curves into a single point for each spacecraft orbit.  Figure \ref{fig:lyman_alpha_lc} shows the resulting light curves for the integrated red and blue wings for all three transit observations.  We calculate a photon noise of 1.4\% for the blue wing during the first orbit and 1.0\% for subsequent orbits; for the red wing, it is 2.8\% for the first orbit and 2.0\% for subsequent orbits.  We conservatively adopt the first-orbit error for all orbits, since instrumental systematics and stellar variability undoubtedly inflate the noise beyond the photon limit.  Using these error bars, we find that the excess absorption during the transit of planet c is $11.1 \pm 1.5$\% in the blue wing and $8 \pm 3$\% in the red wing.  For planet b, we place a 2$\sigma$ upper limit on the in-transit absorption of 3\% in the blue wing and 4\% in the red wing.  Incidentally, the standard deviation of the blue fluxes for all orbits other than the 5 bracketing the transit of c is 1.8\%; the standard deviation of the red fluxes for these same orbits is 3.1\%.  This indicates our inflated error bars of 1.4\% and 2.8\% are not far from the mark.

We illustrate the wavelength dependence of the absorption from planet c by plotting the excess absorption spectrum, $1 - F/F_{\mathrm{out}}$, for each orbit in Figure \ref{fig:excess_spectrum_c}.  Initially, we tried using the post-SAA segment of the observations for the out-of-transit spectrum.  However, we noticed that this introduced significant correlated noise into the excess absorption spectrum, which we attributed to changes in the intrinsic stellar spectrum in the 18 hours between the two segments.  This variability can also be seen in the red wing light curve in Figure \ref{fig:lyman_alpha_lc}, which shows that the red wing is notably dimmer in the post-SAA segment than in the out-of-transit orbits of the pre-SAA segment.  Unfortunately, this variability makes the post-SAA orbits much less useful as a baseline than we had originally hoped.  We considered using the average of the first and fifth orbits for the out of transit spectrum, but the blue wing light curve shows that the fifth orbit might contain planetary absorption.  We therefore opted to use the first orbit for the out of transit spectrum, but note that this first orbit often has an anomalous flux level when compared to later orbits, although it is still a better baseline than the average of the post-SAA spectra.  The resulting plot provides a useful illustration of the progression of the excess absorption spectrum from orbit to orbit, but we should not place too much weight on the absolute value of each spectrum.

Examining Figure \ref{fig:excess_spectrum_c}, we see that the excess absorption is highest in the region of the blue wing near $-10$ km/s (i.e., closest to the line center).  The absorption in this region increases steadily from each orbit to the next until orbit 3, after which it decreases in orbit 4, and decreases again in orbit 5, but does not decrease to 0.  The excess absorption in orbit 3 is $26 \pm 6$\%, more than double the wing-integrated excess absorption of 11\%.  Reassuringly, the excess absorption spectrum decreases blueward of $-10$ km/s until it is indistinguishable from 0 at $-100$ km/s.  It is concerning that the excess absorption in the red wing is highest around 150 km/s, not at the low-velocity edge of 100 km/s where we might expect to see it.  However, the magnitude of the measured absorption in the red line is lower (8\% vs. 11\%) and the noise is much higher, making the data statistically consistent with a flat or slightly declining absorption spectrum in the red wing.

\subsubsection{Stellar variability in Ly$\alpha$}
Since HD 63433 is a young star, it is important to determine whether stellar variability could explain the absorption signal.  Only the visit containing the transit of planet c shows significant blue wing variability, which we are ascribing to planetary absorption.  Among the other 4 visits, which contain a total of 16 orbits, the blue wing is remarkably stable, with a standard deviation of 1.8\%.  The only outlier is the first orbit of the second visit of the successful b observation.  However, the first orbit in a visit is expected to be more variable than the others.  The science exposure is shorter, giving rise to 30\% higher photon noise.  It also starts at a later HST orbital phase, violating our logic for looking at orbit-aggregated data points--namely that repeatable systematics that depend on orbital phase will be averaged out.  Finally, it takes HST approximately one orbit to thermally relax after pointing to a new target, and the first orbit of STIS exoplanet observations at optical wavelengths is routinely discarded because of the higher systematics (e.g. \citealt{huitson_2013,von_essen_2020}).

We can place this level of stellar variability in context by comparing to observations in the literature.  \cite{llama_2016} used disk-resolved Ly$\alpha$ images of the Sun to estimate that for stars with solar activity levels we would expect to see the measured transit depth vary by 0.8\% due to activity-induced measurement error, and are unlikely to see more than 1.5\% variability.  However, HD 63433 is more active than the Sun.  Following \cite{kulow_2014}, we examine observations of the CII line by \cite{loyd_2014}.  The CII line has a formation temperature similar to that of Ly$\alpha$, making it a good tracer of variability in this line.  We compare to observations of Pi UMa, a G1.5 star with a fast rotation period (P=4.89 d) which, like HD 63433, is a member of the Ursa Major Moving Group.  \cite{loyd_2014} found that the mean-normalized excess noise on 60 s timescales for this star was 3.2\%.  For the 28 Myr G1.5 star EK Dra, the excess noise was less than 1\%; for the 13 Myr G1.5 star HII1314, it was less than 5.6\%.  If we assume that the stellar variability has a comparable magnitude on several hour timescales (i.e., the duration of a transit) we might expect HD 63433 to vary by a few percent.  This would suggest that it is unlikely that stellar variability caused the 11\% decrease in brightness in the blue wing during the transit of planet c.

Ly$\alpha$ observations of exoplanet hosts are somewhat less encouraging.  \cite{bourrier_2017c} saw a 20\% dip in Ly$\alpha$ during the transits of sub-Earths Kepler-444e and f, as well as a 40\% dip when no known planet was transiting.  Although Kepler-444 is an old (11 Gyr) K star, the authors couldn't exclude the possibility that the observed variability was due to stellar activity.  \cite{bourrier_2017b} observed HD 97658, an old K star, with STIS over three visits (15 orbits in total) and found that, during the second visit, the Ly$\alpha$ flux declined by 20\% over a period of several hours.  This decline did not coincide with the white light transit and did not have a clear transit-like shape.  It is unclear whether these variations are due to stellar variability or instrumental artifacts.

After considering the totality of the evidence, we conclude that the blue wing absorption from planet c is very likely to be planetary.  It occurs at the expected time and becomes stronger as one approaches the core of the line, in accordance with physical expectations.  In every other visit, the blue wing flux is remarkably stable.  Our HST program will observe a second transit of planet c to see if the signal re-appears, which would provide a definitive confirmation of its planetary origin.  Unfortunately, due to an alignment between c's orbital period and the visibility period imposed by the South Atlantic Anomaly, the next observing window is unlikely to occur before 2023.

The red wing absorption detection is more tentative.  The red wing is much fainter than the blue wing, and has a correspondingly high level of photon noise.  The out-of-transit variability in this wing appears to be higher, and the post-SAA visit for planet c is almost as low as the lowest in-transit data point.  The excess absorption spectrum also appears to rise toward higher velocities, in contravention of theoretical expectations, although the rise is not statistically significant.  On the other hand, the timing and shape of the transit light curve is strikingly similar to that of the blue wing.  We consider it more likely than not that the red wing absorption is real, but without a second transit observation, the detection remains tentative at best.

\subsubsection{A search for absorption in other UV lines}
We can use these same spectra to search for planetary absorption in the Si\,III line at 1206.5\ang and the two N\,V lines at 1238.8\ang and 1242.8\ang.  For c, we see a marginal transit-like signal of $6 \pm 3$\% in Si\,III but no transit-like signal in N\,V ($1.5 \pm 4$\%).  The Si\,III line is known to be highly variable \citep{dos_santos_2019}, and we measure a relative flux of $0.89 \pm 0.02$ for this line in the last orbit, 11\% lower than in the preceding orbits.  For planet b, the Si\,III line is even more variable, and we see no transit-like feature.  We calculate an excess absorption of $1.5 \pm 2.5$\% in Si\,III and $-1.5 \pm 4$\% in N\,V.

\subsubsection{Independent analyses of the Ly$\alpha$ data}
The fiducial analysis reported above was performed by the first author (Michael Zhang, MZ).  Two separate analyses were performed by co-authors Luca Fossati (LF) and Leonardo dos Santos (LDS) using independent pipelines.  There was no communication between the co-authors during these independent analyses other than to agree on a common velocity range in which to look for absorption: [$-140$,$-10$] km/s in the blue wing and [100,200] km/s in the red wing.  The results of these independent analyses are plotted in the Appendix (Section \ref{section:appendix_alt_analyses}).  All three analyses show a clear blue wing absorption signal from c, and no red or blue wing absorption from b.  The alternative analyses show no red wing absorption from c.  This is likely due to their 3x higher scatter (see Appendix), but the non-detection of red wing absorption in these alternative analyses nevertheless underscores the tentative nature of the detection in the fiducial analysis.

\subsection{Helium absorption during transit}
\begin{figure}
  \centering 
  \subfigure {\includegraphics
    [width=0.5\textwidth]{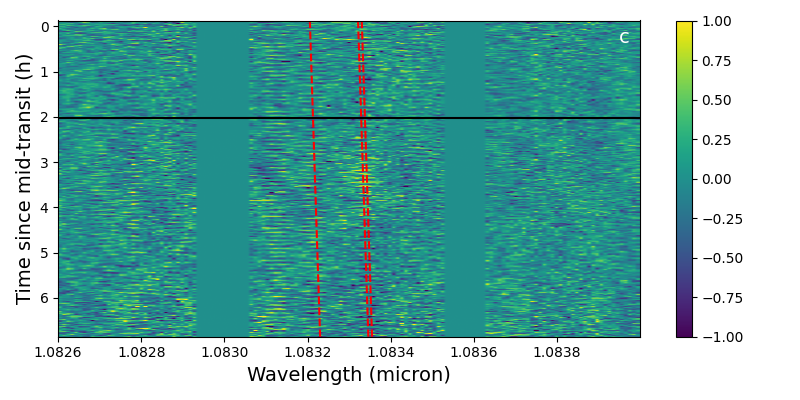}}
   \subfigure {\includegraphics
    [width=0.5\textwidth]{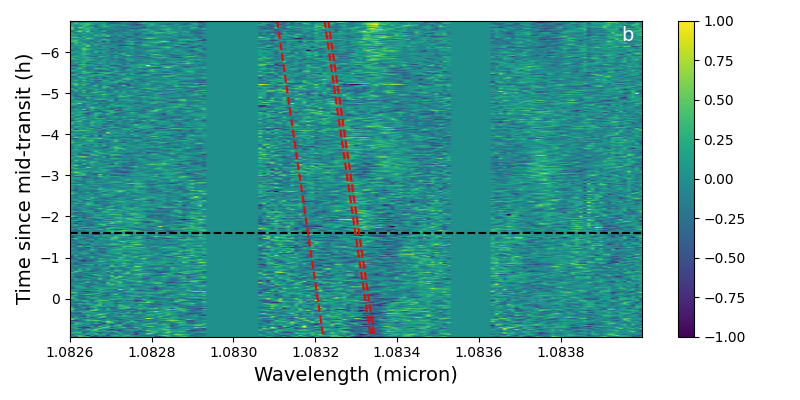}}
    \caption{Residuals images after SYSREM, for planet c (top) and b (bottom).  Colors indicate relative brightness change in percent.  The dashed red lines indicate the wavelengths of the helium triplet in the planetary frame.  The dashed horizontal black line indicates the beginning of transit, while the solid horizontal black line indicates the end of transit.  Note the darkening during the transit of b, which we ascribe to stellar variability (see text).}
\label{fig:after_sysrem}
\end{figure}

\begin{figure}
  \centering 
  \subfigure {\includegraphics
    [width=0.5\textwidth]{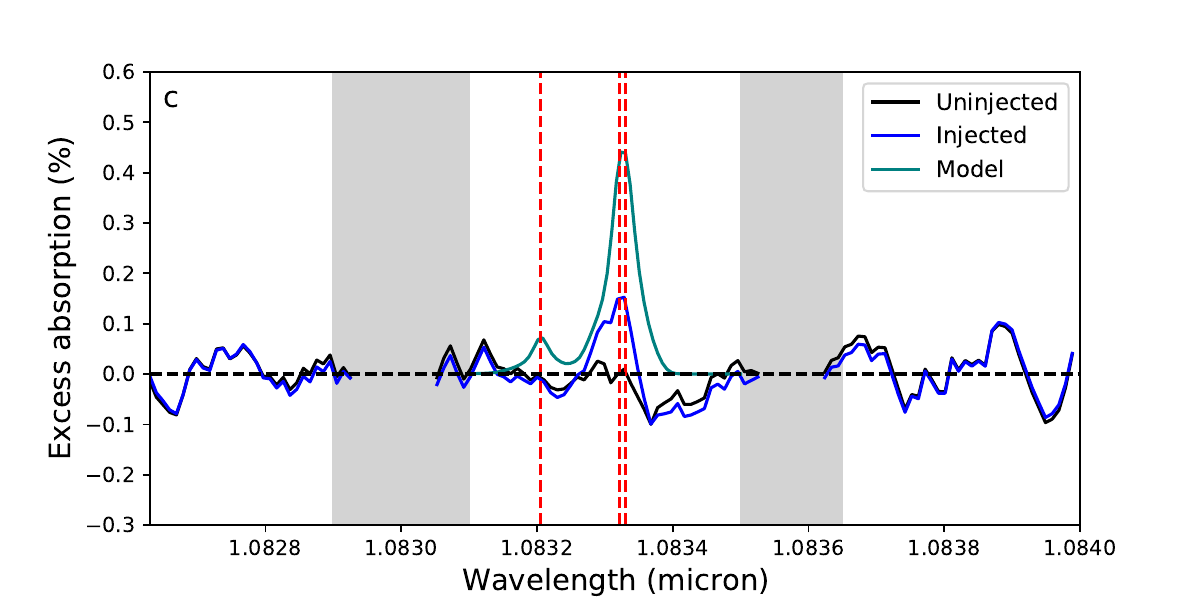}}
   \subfigure {\includegraphics
    [width=0.5\textwidth]{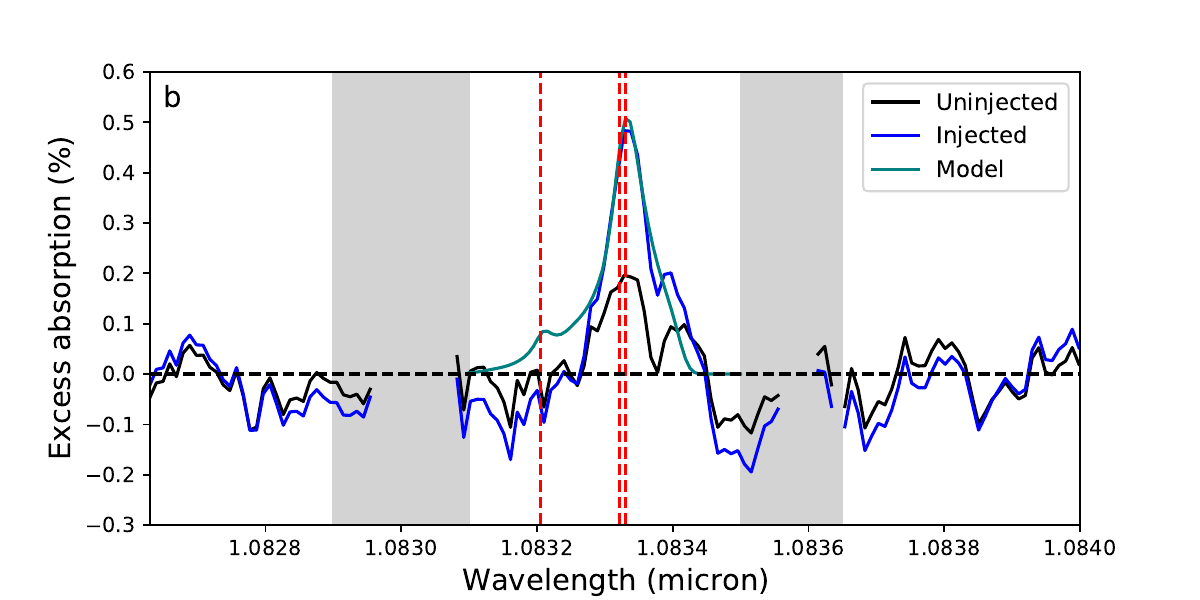}}
    \caption{In black is the excess absorption spectrum in the planetary frame for c (top) and b (bottom).  The vertical red lines indicate the positions of the helium triplet.  The absorption from planet b is probably stellar variability (see text, and Figure \ref{fig:after_sysrem}).  The injected models (teal) are from the 3D hydrodynamic simulations described in Section \ref{sec:modeling}, and the observed excess absorption spectrum after injection is shown in blue.  The grey regions are locations of strong stellar and telluric lines.}
\label{fig:helium_excess_planetary}
\end{figure}

We next examine the Keck data to search for signs of helium absorption during the transits of planets b and c.  After the processing steps described in Subsection \ref{subsec:keck_reduction}, we are left with a residuals image of size N\textsubscript{epochs} by N\textsubscript{wavelengths}.  Each pixel in the residuals image approximately represents the fractional flux change at that epoch and wavelength from the mean spectrum.  We shift these spectra to the planetary frame, combine all out-of-transit spectra into a master out-of-transit spectrum, combine all in-transit spectra to a master in-transit spectrum, and subtract the master in-transit spectrum from the master out-of-transit spectrum.  Figure \ref{fig:helium_excess_planetary} shows the resulting excess absorption spectrum for each planet.

\begin{figure}
    \centering
    \subfigure {\includegraphics
    [width=0.5\textwidth]{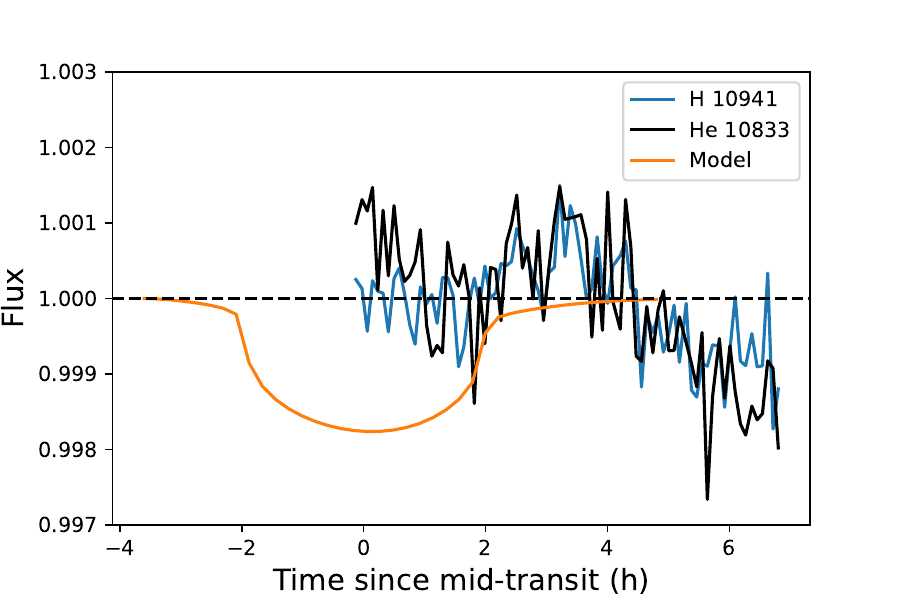}}
   \subfigure {\includegraphics
    [width=0.5\textwidth]{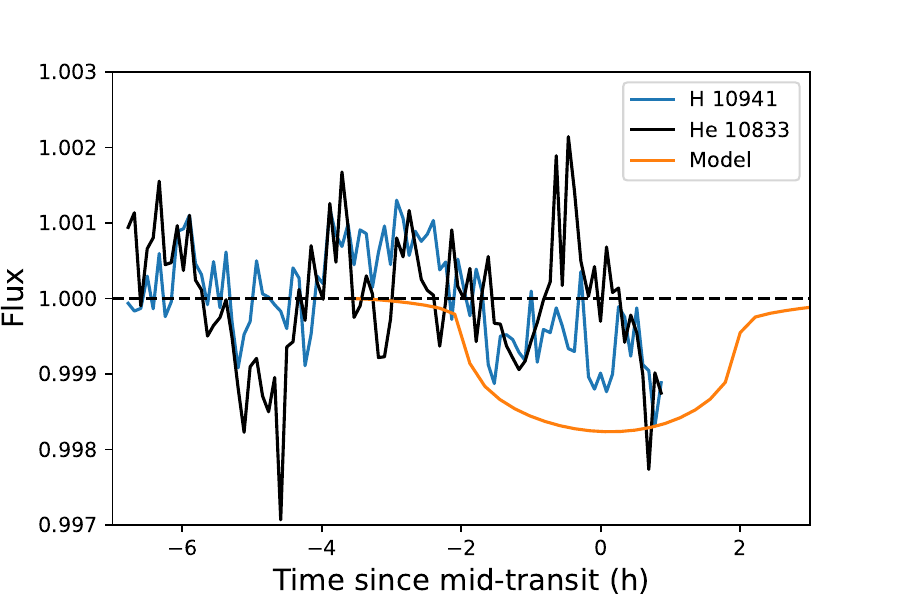}}
    \caption{Light curves of the HeI 10833 \AA{} and HI 10941 \AA{} lines, integrated in a 3.2 \AA{} bandpass, for planet c (top) and b (bottom).  To normalize out non-stellar variations, the line fluxes are divided by the continuum at 10825 \AA{} (HeI) or 10937 \AA{} (HI), also integrated in a 3.2 \AA{} bandpass.  In orange we plot the predictions of the 3D hydrodynamic model.}
    \label{fig:lines}
\end{figure}

We see clear evidence of stellar activity in Figure \ref{fig:after_sysrem}, and in the line-integrated line fluxes (Figure \ref{fig:lines}).  The stellar 10833 \ang helium lines, which trace chromospheric activity, are variable on both nights.  During the transit of planet c on the first night, the lines experience a bump in brightness starting around 3 hours after mid-transit and then fall to even lower values after the bump.  During the second night, the helium lines exhibit a more complicated behavior: they start high, then decline for 2 hours, rise again, decline again, rise a third time, and finally decline to their lowest level over the night.  The peak-to-peak amplitude of this variability is around 1\% on the first night and 2\% on the second night.

We conclude that the 0.2\% excess absorption signal in the helium lines for planet c (Figure \ref{fig:helium_excess_planetary}) is most likely due to stellar variability, not planetary absorption.  It is apparent in Figure \ref{fig:after_sysrem} that the excess absorption signal is caused by the darkening of the stellar spectrum around the 10833\ang helium lines in the final hour of the observations.  A close examination of the darkened portion of the spectrum shows that the darkening started, not at the beginning of the transit, but an hour afterwards.  In addition, the darkened portion of the spectrum does not follow the radial velocity of the planet.  On the contrary, it moves toward shorter wavelengths as time progresses.  This is the opposite of what we would expect a planetary absorption signal to do.  Finally, 0.2\% is smaller than the observed amplitude of the variability in the stellar helium lines on both nights.  Although we cannot rule out a planetary or hybrid planetary-and-stellar explanation for the observed absorption signal, the data are entirely consistent with a purely stellar explanation.

We next explore whether or not we might be able to model out some of this stellar variability using other chromospheric lines in our spectrum.  The most promising candidates are two lines in the hydrogen Paschen series: $\gamma$ (n=$6 \rightarrow 3$, 10941 \ang) and $\delta$ (n=$7 \rightarrow 3$, 10052 \ang).  We use the data from the second night, which had more complex stellar behavior than the first, as our test case.  We find that the $\gamma$ line has a time-varying behavior similar to that of the helium lines (Figure \ref{fig:lines}), but with a lower overall amplitude.  The $\delta$ line is also variable, but this variability does not appear to be correlated with the variability in the helium lines.  Ultimately, neither line displayed a strong enough correlation with the helium lines to enable an effective correction for stellar activity.  However, the similar behaviors of the Paschen $\gamma$ and helium lines provide additional support for our conclusion that the variability in Figure \ref{fig:after_sysrem} is likely stellar and not planetary in origin. 

We next consider what limits we can place on the magnitude of helium absorption during the transits of planets b and c.  Planet c barely accelerates during its transit, making it hard to disentangle planetary signals from the stellar and telluric variability that SYSREM is meant to subtract.  As a result, we expect significant self-subtraction from our analysis pipeline.  Planet b accelerates more and should experience less self-subtraction.  Figure \ref{fig:helium_excess_planetary} illustrates this phenomenon: after injecting an artificial helium absorption signal, our measured excess absorption spectrum is only 35\% the size of the injected signal (i.e., 65\% self-subtraction).  For planet b, the injection-recovery test indicates a self-subtraction of 40\%.

Due to stellar variability, we cannot assume statistical independence between epochs and use the standard statistical methods to compute an upper limit on the helium excess absorption.  We can, however, arrive at a reasonable guess by examining the observed stellar variability during each of the two nights and its corresponding effect on the excess absorption spectrum.  The helium lines never deviate by more than 1\% from the median on either night, and even if they did, it is unlikely that the stellar variability would line up with the planetary transit.  For planet b, the acceleration of the planet is significant enough to place the planetary absorption lines outside of the stellar lines at the beginning of the transit, further decreasing the impact of stellar variability.  A 1\% planetary absorption would be reduced to 0.5\% due to self-subtraction, but this would still be readily detectable in the excess absorption spectrum in Figure \ref{fig:helium_excess_planetary}.  We test this by injecting a planetary signal with an amplitude of 1\% into the data, running it through the pipeline, and examining the intermediate outputs.  Even in the pre-SYSREM stage (before any self-subtraction happens), the planetary signal is clearly visible above the amplitude of the stellar variability.  The planetary signal remains obvious when we reduce the peak excess absorption of the injected signal to 0.5\%, but its final amplitude in that case would be comparable to the amplitude of the stellar variability during the transit of planet c.  We therefore conclude that the peak excess absorption must be less than 1\% for both planets with high confidence, and less than 0.5\% with medium confidence.

\section{Understanding the star}
\label{sec:understanding_star}
In order to model mass loss, we need to know the star's intrinsic spectrum at high energies.  Heating from the X-ray and extreme UV (EUV) flux drives the outflow in our models, while the stellar Ly$\alpha$ line profile, in combination with the instrumental line spread profile, is necessary to predict the observed absorption using the models.  For the star's X-ray spectrum, we use the XMM-Newton data described in \S\ref{subsec:xmm_data_reduction}, and combine these data with older observations at X-ray and optical wavelengths to characterize the star's long-term variability and activity cycle.  For the star's UV spectrum, we use a combination of scaling relations and, when appropriate, observations of the Sun's UV spectrum.  

\subsection{X-ray spectrum and stellar variability}
\subsubsection{XMM-Newton spectrum}
We analyze the XMM-Newton EPIC observations using the \texttt{xspec} package.  To get the underlying X-ray spectrum, we fit a model equal to the sum of two APEC emission models.  These models \citep{smith_2001} assume an optically thin, collisionally ionized plasma with the temperature, metallicity, redshift, and normalization as free parameters.  We fix the redshift to 0 (EPIC's velocity resolution is $>$6000 km/s), require the two model components to have the same metallicity, and let the two temperatures, two normalization factors, and the global metallicity vary freely.  We fit the data by minimizing the W statistic, the analogue of $\chi^2$ for a distribution corresponding to the difference of two Poisson distributions (namely source and background).

\begin{figure}
  \centering 
  \subfigure {\includegraphics
    [width=0.5\textwidth]{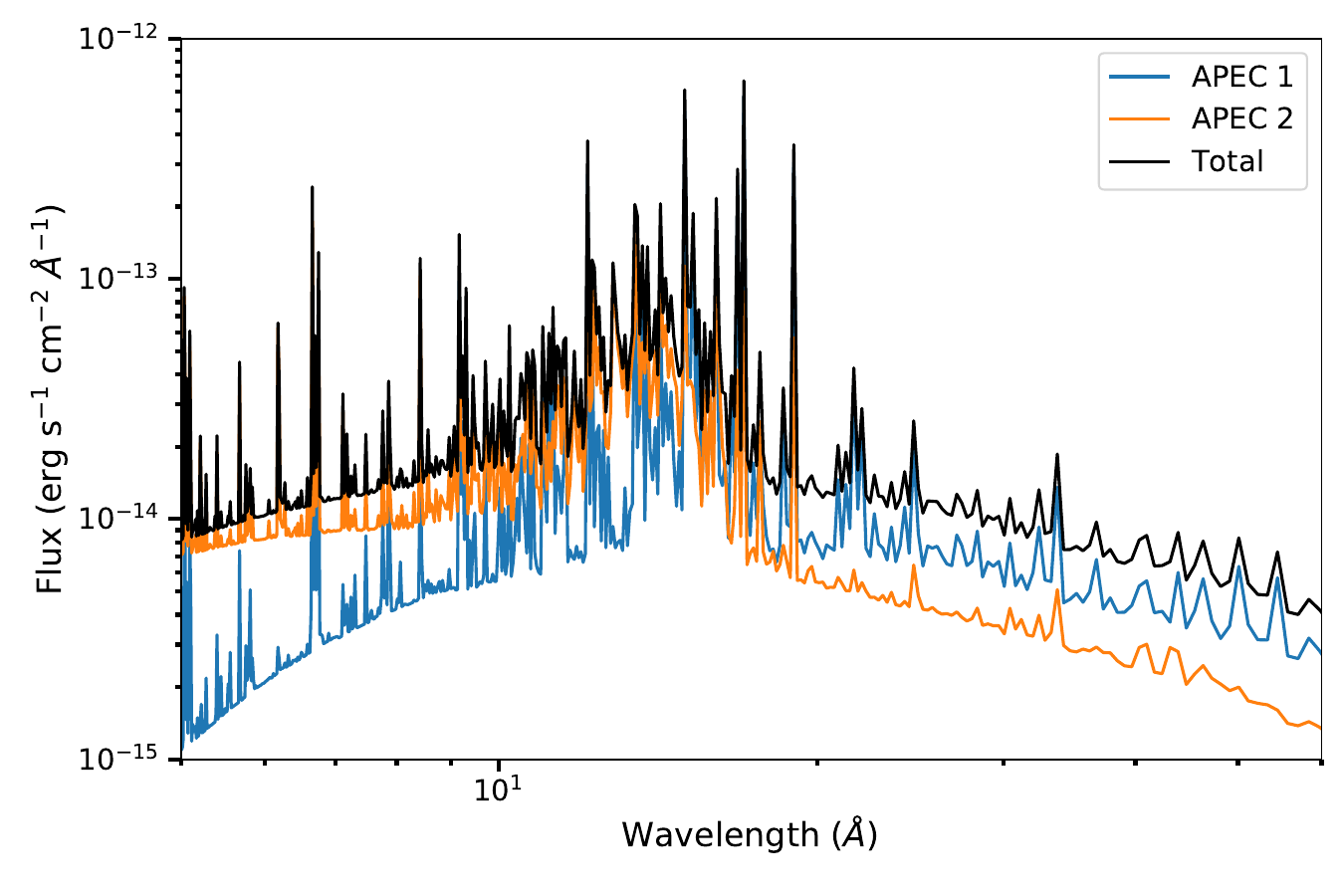}}
   \subfigure {\includegraphics
    [width=0.5\textwidth]{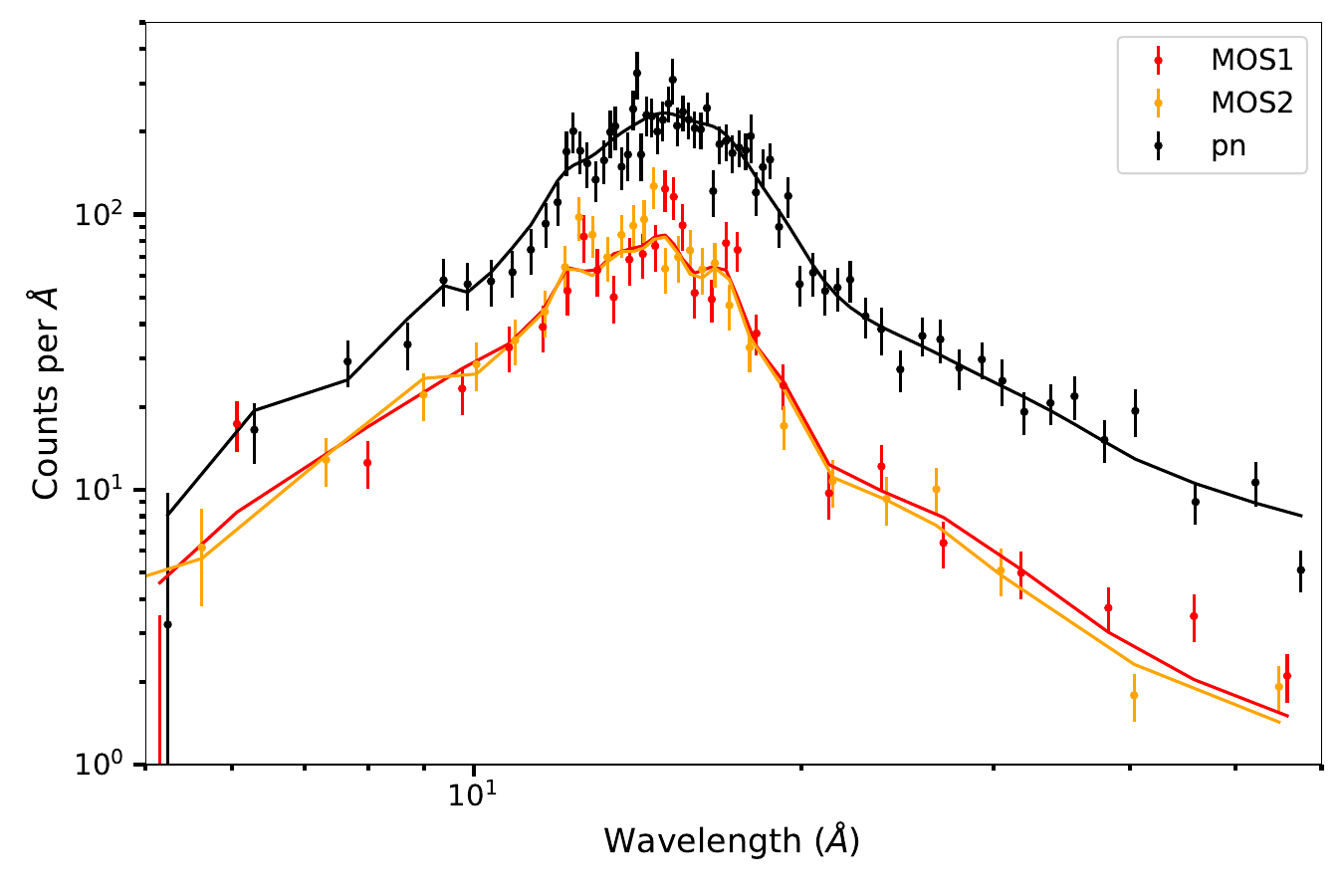}}
    \caption{Top: Best-fit intrinsic stellar X-ray spectrum.  The total spectrum as observed from Earth (black) is the sum of a low-temperature component (APEC 1) and a high-temperature component (APEC 2).  Bottom: the XMM EPIC spectra measured by the three detectors (MOS1, MOS2, pn) and the folded models, which take into account the instrumental RMF and ARF (the X-ray equivalents of LSF and throughput).  For clarity, the data are binned so that each bin, except those on the edges, contains at least a 3$\sigma$ detection.}
\label{fig:xmm_epic}
\end{figure}

\begin{table}[ht]
  \centering
  \caption{Model parameters for XMM-Newton data}
  \begin{tabular}{c C}
  \hline
  	  Parameter & \text{Value}\\
      \hline
      Metallicity & 0.44 \pm 0.07\\
      kT$_1$ (keV) & 0.38 \pm 0.02\\
      EM$_1$ (cm$^{-3}$) & 3.8 \pm 0.5 \times 10^{51}\\
      kT$_2$ (keV) & 0.83 \pm 0.04\\
      EM$_2$ (cm$^{-3}$) & 2.1_{-0.4}^{+0.3} \times 10^{51}\\
      Flux$^*$ (erg/s/cm$^2$) & 1.25_{-0.02}^{+0.03} \times 10^{-12}\\
      \hline
  \end{tabular}
  \tablecomments{$^*$Derived, not a fit parameter.  For the range 5-100 \AA{} (0.124-2.48 keV).}
  \label{table:xmm_params}
\end{table}

Figure \ref{fig:xmm_epic} shows the EPIC data, the best fit to the data obtained by \texttt{xspec}, and the intrinsic spectrum implied by the best fit parameters.  We ran a Markov Chain Monte Carlo (MCMC) fit using the Metropolis-Hastings algorithm and a chain length of 10,000 to estimate the range of parameters consistent with the data.  We plotted the chain to ensure convergence, which occurred within the first 1000 samples.  Table \ref{table:xmm_params} shows the resulting 1D MCMC posteriors.  The metallicity is reported with respect to the solar abundances of \cite{asplund_2009}.  We find a dominant component with a temperature of 0.38 keV (4.4 MK), with a slightly sub-dominant component at 0.83 keV (9.6 MK).  The emission measures are on the high end compared to other moderately active G8-K5 dwarfs \citep{wood_2010}, but of the same order of magnitude.  The best-fit metallicity is subsolar, and because the photospheric metallicity of the star is roughly solar ([M/H] = $-0.09 \pm 0.08$; \citealt{mann_2020}), it is substellar as well.  This is reminiscent of the findings of \cite{poppenhaeger_2013}, who observed the moderately active K dwarf HD 189733 A with Chandra.  In that study they fit the O, Ne, and Fe abundances separately, obtaining values of $0.31 \pm 0.02$, $0.25 \pm 0.13$, and $0.64 \pm 0.05$ (relative to solar), respectively.  We carried out an analogous fit where we allowed the O, Ne, and Fe abundances to vary freely and obtained $0.67 \pm 0.1$, $0.22_{-0.05}^{+0.40}$, and $0.97_{-0.18}^{+0.05}$, respectively.  These results are due to an effect called the first ionization potential (FIP) bias: for many inactive and moderately active stars, elements with high FIP (e.g. C, O, N, Ne) are depleted in the corona compared to low FIP elements (e.g. Mg, Si, Fe).  FIP bias was seen by \cite{wood_2010} in 5 out of 7 moderately active G8-K5 dwarfs.  In these stars, the coronal abundances of C, O, N, and Ne were all lower than the photospheric abundances.  FIP bias is also seen in the solar corona, although there, low FIP elements are enhanced by a factor of $\sim3$ and high FIP elements generally have photospheric abundances \citep{feldman_2002}.  It has been suggested that the FIP bias may arise from wave ponderomotive forces on the upper chromosphere \citep{laming_2004,laming_2017}.  For our purposes here, it is sufficient to reconstruct the intrinsic X-ray spectrum of the host star and we therefore leave further analysis of the FIP bias to interested stellar astronomers.

\begin{figure}
    \includegraphics[width=0.5\textwidth]{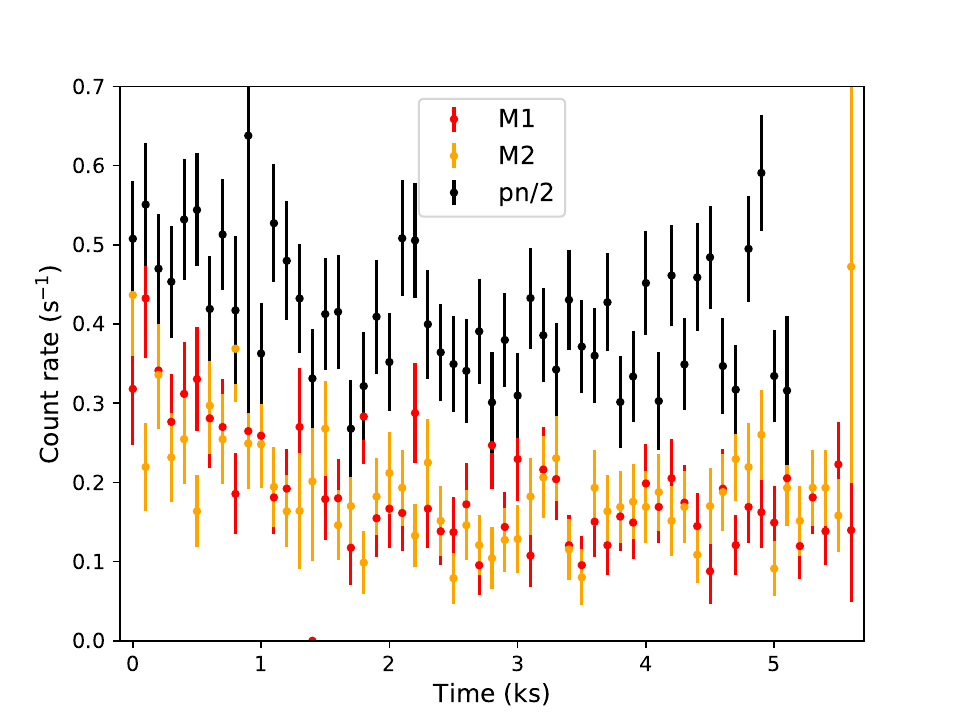}
    \caption{Background-subtracted X-ray light curves of HD 63433 recorded by the three EPIC cameras (MOS1, MOS2, pn).  The pn light curve is divided by 2 for clarity.}
    \label{fig:xmm_epic_lc}
\end{figure}

Our fit to the EPIC data constrains the time-averaged 5--100 \AA{} (0.124-2.48 keV) flux to $1.25_{-0.02}^{+0.03} \times 10^{-12}$ erg/s/cm$^2$.  Restricting the range to the observable range of both XMM and ROSAT, we find a 0.2--2.4 keV flux of $1.12 \pm 0.02 \times 10^{-12}$ erg/s/cm$^2$.  These error bars are deceptively small, as the X-ray spectra of active stars can vary significantly in time.  Figure \ref{fig:xmm_epic_lc} shows the X-ray light curve captured by the three EPIC cameras.  HD 63433 is brighter in the first 2 ks of observation than in the remaining 4 ks, with the pn-flux declining by 25\% and the two MOS fluxes declining by 50\%.  The difference in observed amplitude is likely due to the different characteristics of the two types of detectors.  The MOS detectors' sensitivities drop off more sharply toward low energies ($<$1 keV), where most of the star's X-ray flux resides, than the pn-detector.  The observed variability in the X-ray light curve means that, absent simultaneous observations, we cannot know the X-ray flux during the epoch of our hydrogen or helium observations to better than $\sim$25\%.

\subsubsection{Long-term X-ray variability from ROSAT data}
We evaluate the magnitude of the stellar X-ray variability over longer timescales using archival ROSAT data from 1990. These data have a much lower SNR than the XMM-Newton data, as a result of the lower effective area of the detector and the shorter exposure time.  Whereas XMM's EPIC cameras captured 3600 X-ray photons, ROSAT's PSPC-C captured only 86.  As with XMM, we fit the data with two summed APEC emission models.  In order to prevent the fit from wandering off to unphysical parts of parameter space, we fix the metallicity to the value derived from XMM and constrain $kT_1$ to lie between 0 and 0.5 keV and $kT_2$ to lie between 0.5 and 1.0 keV.

We find that the shape of the unfolded ROSAT spectrum is consistent with the shape of the unfolded XMM spectrum.  We derive a 0.124--2.48 keV flux of 1.5--1.9 $\times 10^{-12}$ erg/s/cm$^2$ and a 0.2--2.4 keV flux of 1.25--1.71 $\times 10^{-12}$ erg/s/cm$^2$.  This is in line with the flux reported by the Second ROSAT all-sky survey (2RXS) source catalog \citep{boller_2016} for both a power-law fit (1.39 $\times 10^{-12}$ erg/s/cm$^2$) and a blackbody fit (1.77 $\times 10^{-12}$ erg/s/cm$^2$).    We conclude that the star's X-ray flux appears to have been 33\% higher during the ROSAT observation than during the XMM observation, but the two measurements are consistent at the $2\sigma$ level.

\subsubsection{Long-term optical variability from the APT data}
\begin{figure}
    \includegraphics[width=0.5\textwidth]{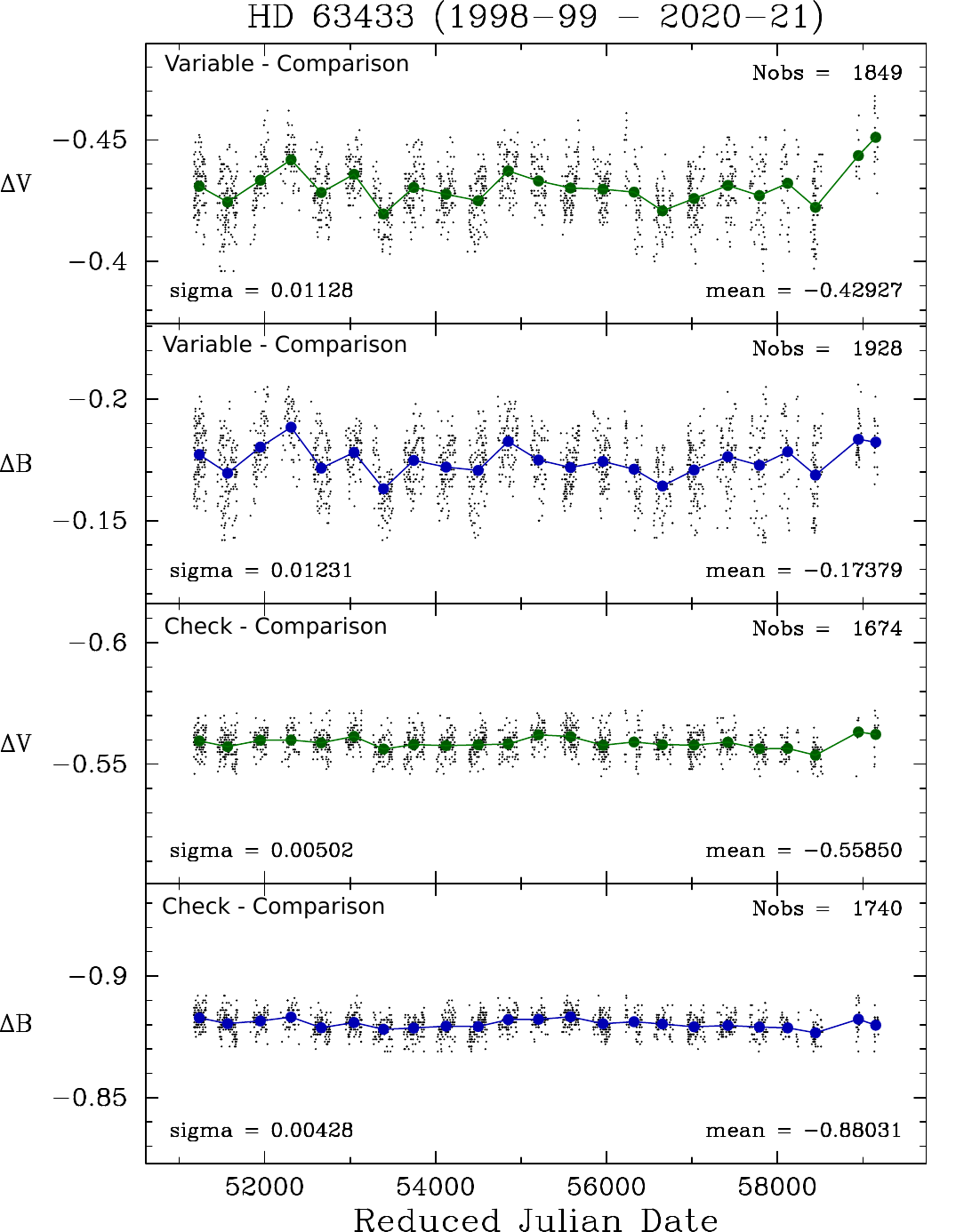}
    \caption{Light curve for HD 63433 (`Variable') in the Johnson V (green; top panel) and B (blue; upper middle) pass bands normalized using a comparison star.  Nightly observations are plotted as small grey points, while the annual means are shown as larger colored circles.  An equivalent light curve for a less active star, HD 63432 (`Check'), is shown in the lower middle and bottom panels to demonstrate the photometric stability of these data.}
    \label{fig:apt}
\end{figure}

We use ground-based optical photometric monitoring data to evaluate where HD 63433 was in its activity cycle during the epochs of our HST and Keck observations.  HD 63433 has been monitored since 1998 in the Johnson V and B photometric pass bands with the Tennessee State University T3 0.40 m Automatic Photoelectric Telescope (APT) at Fairborn Observatory in southern Arizona.  Our observations cover 23 observing seasons from 1998-99 to 2020-21, although there are relatively few observations in the last three seasons due to APT scheduling changes and instrument problems.  While the typical season runs from early October to late April, the 2019-2020 data only span March 15, 2020 to April 19, 2020, and the 2020-2021 data only span October 12, 2020 to November 25, 2020.

Our measurements of HD 63433 were made differentially with respect to the comparison star HD 64465 (HIP 38677; F5). A second star, HD 63432 (HIP 38231; A2), was used to check the stability of the relative photometry using HD 64465.  Details of the robotic telescopes and photometers, observing procedures, and data reduction can be found in \citet{henry_1999} and \citet{fekel_2005}.  Figure \ref{fig:apt} shows the Variable minus Comparison and Check minus Comparison APT light curves.  The lower two panels show small variability in the Chk-Cmp light curves.  Their seasonal means vary over a range of 0.006 and 0.008 mag in the V and B, respectively.  However, the Var-Cmp seasonal means in the upper two panels vary over a much larger range of 0.025 and 0.022 mag, demonstrating that most of the variability seen in the Var-Cmp light curves is intrinsic to HD 63433.

Keeping the data limitations in mind, the APT light curve does appear to show that HD 63433 was anomalously bright in 2020 and 2021 in both filters, with a flux 1-2\% above the 20 year average in V.  In fact, it appears to be brighter than at any point since observations began.  For active stars ($\log{R'_{\mathrm{HK}}}>-4.7$), the V and B brightness varies inversely with stellar activity: the more active the star, the more spots it has, and the dimmer it appears \citep{lockwood_2007}.  Therefore, the APT data suggest that HD 63433 was unusually quiescent during the time of our mass-loss observations, consistent with the marginally lower X-ray flux observed by XMM-Newton in March 2021 compared to ROSAT in 1990.

We computed a Lomb-Scargle Periodogram of the data and found a clear, narrow peak corresponding to the rotation period of the star.  Both the V and B data indicate $P_{\rm rot}=6.413 \pm 0.002$ d, slightly lower than the $6.46 \pm 0.01$ d reported by \cite{gaidos_2000} using the APT data available then.  The APT-inferred rotation period is far more precise than the one inferred from TESS and K2 data \citep{mann_2020} because of the much longer baseline (1200 rotation periods vs. 3-4). The periodogram also shows a broad double peak around 800--1100 d, perhaps indicative of a short stellar cycle.  However, while there is clear non-random behavior in the light curve on a timescale of years, no periodic stellar cycle is obvious by inspection.  HD 63433's photometric variability is typical of stars younger than 2--3 Gyr, which have complex interannual variations that are often composed of multiple cycles, compared to the simple cycles of older stars \citep{olah_2016}.

\subsection{UV spectrum}
\subsubsection{Ly$\alpha$ profile}
In order to translate our mass loss models into a prediction for the Ly$\alpha$ light curve during transit, we need a measurement of the star's intrinsic Ly$\alpha$ profile.  The STIS observations do not directly tell us the intrinsic Ly$\alpha$ profile because the the line core is absorbed by the ISM, and the instrumental line-spread profile smears out the remaining flux.  We reconstruct the intrinsic profile from our data using a hierarchical Bayesian model implemented in \texttt{stan} \citep{stan2018}.

In principle there are an infinite number of intrinsic profiles that can fit the data, because the flux at the core of the line is unconstrained.  Therefore, we need to utilize a prior on the line shape in order to reconstruct the height of the line core using the flux in the wings.  In a previous survey of stellar Ly$\alpha$ emission, \cite{wood_2005} used the profile of the observed Mg II h and k lines (2796 and 2804 \AA{}) as their template for the Ly$\alpha$ line shape.  Unfortunately, we have no such data for HD 63433.  Instead, we start with the reconstructed Ly$\alpha$ profile of HD 165185 from \cite{wood_2005}, a star with the same spectral type as HD 63433 and a similar rotation period.  We then allow \texttt{stan} to modify the profile as follows:

\begin{align}
    F(\lambda) = F_{\rm init}(\lambda) + \Delta_1(\lambda)\\
    \Delta_1(\lambda) = \text{cumsum}(\Delta_2(\lambda))\\
    \Delta_2(\lambda) \sim N(0, 5 \times 10^{-14}),
\end{align}

where $F_{\rm init}$ is the HD 165185 profile, cumsum is the cumulative sum, and N(0, $5 \times 10^{-14}$) is a normal distribution with a mean of 0 and standard deviation of $5 \times 10^{-14}$.  The intrinsic spectrum of HD 63433 is that of HD 165185 plus differences, the differences are in turn the cumulative sum of second differences, and we impose a Gaussian prior on the second differences with a standard deviation of $5 \times 10^{-14}$ erg s$^{-1}$ cm$^{-2}$ \AA$^{-1}$.  These equations allow \texttt{stan} to modify the HD 165185 profile to fit the HD 63433 data, but not arbitrarily: it enforces continuity in the modifications made to the profile, and penalizes large changes to avoid overfitting. This process is mathematically equivalent to L2 regularization.

We can obtain an independent constraint on the magnitude of the interstellar absorption by using the parameters derived by \cite{dring_1997} for two stars.  $\beta$ Gem and $\sigma$ Gem are 1.3 and 2.2 degrees, respectively, from HD 63433.  Despite having very different distances (10.3 pc and 37.5 pc), the two have indistinguishable N(HI) of $10^{18.26}$ and $10^{18.20}$ cm$^{-2}$.  This is because the region within 10 pc has an abnormally high neutral hydrogen fraction compared to the rest of the 100 pc Local Bubble \citep{wood_2005}.  \cite{dring_1997} found that the sightline to these stars can be modelled by assuming two clouds: one at 21.7 km/s with a column density of $10^{18.027}$ cm$^{-2}$ and a HI Doppler parameter of 12.35 km/s, and another at 32.5 km/s with a column density of $10^{17.801}$ cm$^{-2}$ and a Doppler parameter of 11.0 km/s.  (In practice, \cite{dring_1997} fit the two sightlines separately, but we averaged the results here because they are remarkably similar.) The more strongly absorbing cloud has a velocity consistent with theoretical expectations.  The local interstellar cloud (LIC) is moving at 25.7 km/s in the direction of l=186$\degree$, b=-16$\degree$, according to high-resolution observations of local stars \citep{lallement_1995}, in good agreement with the flow of ISM particles through the solar system ([l=183$\degree$, b=-16$\degree$] at 26.3 km/s) \citep{witte_2004}.  Projecting this velocity along the line of sight, we compute a radial velocity of 19.9 km/s.  Combined with the -16 km/s radial velocity of the star, the 36 km/s difference is what strongly suppresses the red wing while keeping the blue wing unusually intact.

\begin{figure}
  \centering 
  \subfigure {\includegraphics
    [width=0.5\textwidth]{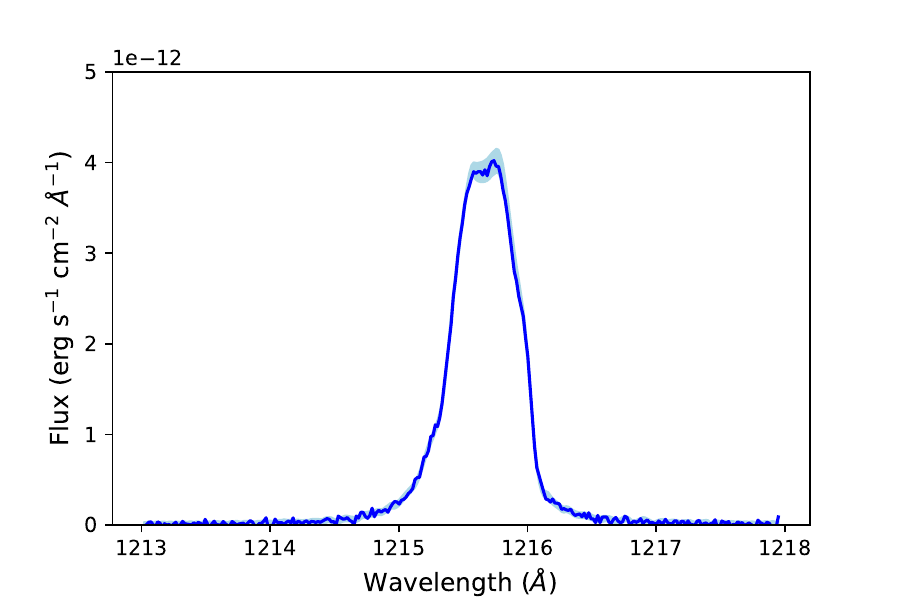}}
   \subfigure {\includegraphics
    [width=0.5\textwidth]{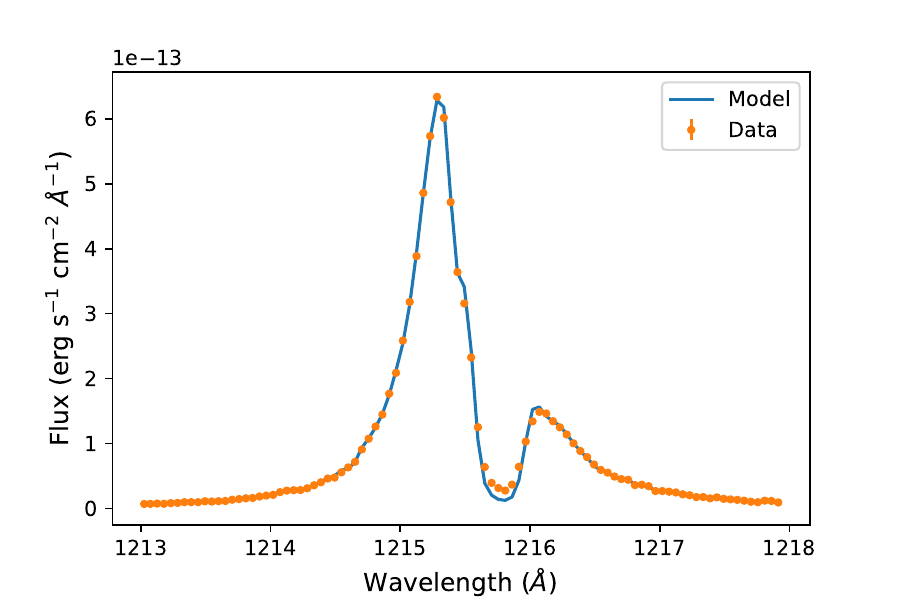}}
    \caption{Top: the reconstructed stellar Ly$\alpha$ profile (blue), with uncertainty indicated by the gray shading.  Bottom: the model Ly$\alpha$ profile after interstellar absorption and convolution with the instrumental line spread profile (blue), compared to the data (orange).  The data has error bars, but they are too small to see.}
\label{fig:ly_spectrum}
\end{figure}

To summarize, our free parameters are the 282 second differences $\Delta_2(\lambda)$, while the interstellar absorption is fixed to that found by \cite{dring_1997}.  We run 10,000 iterations of 4 chains each, and check all five of the diagnostics provided by \texttt{stan} to ensure convergence: effective sample size, potential scale reduction factors, divergent transitions, percentage of transitions ending prematurely due to maximum tree depth, and E-BFMI (energy).  We take the sample with the highest posterior probability to generate the fiducial Ly$\alpha$ profile.  Figure \ref{fig:ly_spectrum} shows the resulting best-fit intrinsic Ly$\alpha$ profile.  The model provides a close match to the data everywhere except at the center of ISM absorption, where the data are higher than the model.  We speculate that this may be because the LSF provided by STScI does not have sufficiently strong wings.  This was previously noted by \cite{bourrier_2017b}, who fit their own LSF for the 52 x 0.05 arcsec slit.

We evaluate how sensitive the Ly$\alpha$ flux is to our choice of reconstruction method by repeating our analysis using a completely independent method similar to the one in \cite{bourrier_2017b}.  In this version, we model the stellar Ly$\alpha$ line as a Voigt profile, the instrumental line spread profile as a sum of two Gaussians, and the interstellar medium as a cloud with a single Gaussian velocity dispersion and velocity.  We then use differential evolution to optimize the free parameters: the $\mu_*$, $\sigma_*$, $\gamma_*$, and amplitude of the stellar line profile, the two standard deviations and one relative amplitude which characterize the LSF, and the velocity offset of the intervening cloud.  The column density of the cloud is set to the sum of the column densities of the two clouds found by \cite{dring_1997}, and the Doppler broadening parameter is set to 12 km/s, very close to the values derived by \cite{dring_1997} for both clouds.  With this method, we obtain a Ly$\alpha$ flux of 46 erg cm$^{-2}$ s$^{-1}$ at 1 AU, 18\% lower than the fiducial value.  Based on the fit error and \cite{linsky_2014}, the Ly$\alpha$ flux we derive is probably not accurate to better than $\sim$30\%.  The fit parameters are given in Table \ref{table:alt_fit_params}.

\begin{table}[ht]
  \centering
  \caption{Parameters derived from alternate fit to Ly$\alpha$ data}
  \begin{tabular}{c c}
  \hline
  	  Parameter & Value\\
      \hline
      $\mu_*$ & 1215.595\ang\\
      $\sigma_*$ & 0.198\ang\\
      $\gamma_*$ & 0.072\ang\\
      $A_*$ & $3.40 \times 10^{-12}$ erg cm$^{-2}$ s$^{-1}$ \AA$^{-1}$ \\
      $\sigma_{\rm lsf,1}$ & 0.18\ang\\
      $\sigma_{\rm lsf,2}$ & 0.10\ang\\
      $A_{\rm lsf,2}/A_{\rm lsf,1}$ & 0.26\\
      $v_{\rm cloud}$ & 16.8 km/s\\
      \hline
  \end{tabular}
  \label{table:alt_fit_params}
\end{table}

\subsubsection{Broader UV spectrum}
We first estimate the shape of the stellar spectrum in the extreme ultraviolet: wavelengths shortward of the Ly$\alpha$ line but longward of 100\AA{}.  These photons ionize hydrogen, which deposits heat into the atmosphere and drives the outflow.  There are currently no EUV telescopes, so we rely on the scaling relations obtained by \cite{linsky_2014} to estimate the EUV flux in 100 \AA~bins between 100--1170 \AA based on the star's measured Ly$\alpha$ flux.  For the wavelength range 100--400 \AA, these scaling relations were based on stellar observations with Extreme Ultraviolet Explorer \citep{bowyer_1994} and Far Ultraviolet Spectroscopic Explorer \citep{sembach_1999}.  For the wavelength range 400--1170 \AA, they are based on the semi-empirical solar model of \cite{fontenla_2014}.

Moving to longer wavelengths, the stellar flux between the Ly$\alpha$ line at 1216 \AA{} and the triplet helium ionization limit of 2588 \AA{} has special importance for helium observations.  These photons are not energetic enough to ionize hydrogen or ground state helium, and therefore do not create the ions or electrons which, through recombination, create triplet state helium.  However, they are energetic enough to destroy triplet ground state helium via ionization.  This means that the level population of triplet helium, and the magnitude of the corresponding triplet helium absorption signal during transit, is sensitive to the value of the stellar flux in this wavelength range.

HD 63433 is a solar analogue, so we adopt the solar MUV spectrum as measured by the Solar Radiation and Climate Experiment (SORCE) satellite.\footnote{https://lasp.colorado.edu/home/sorce/data/} We verify the applicability of this spectrum by comparing to data from XMM-Newton's Optical Monitor, which observed the star's MUV flux through two filters: UVM2 ($\lambda=231 \pm 48$ nm) and UVW2 ($\lambda=212 \pm 50$ nm).  The OM measured $90.2 \pm 0.3 s^{-1}$ in UWM2 and $35.02 \pm 0.15 s^{-1}$ in UVW2.  After correction for coincidence losses due to multiple photons hitting the detector in the same frame, we obtain $128.6 \pm 0.6 s^{-1}$ and $40.2 \pm 0.2 s^{-1}$, respectively.  We used the solar spectrum and the filter transmission profiles to predict what OM would have seen if HD 63433 were an exact solar clone and obtained a count rate within 2\% of the observed rate for UVM2 and within 7\% for UVW2.  We therefore conclude that the Sun's MUV spectrum accurately matches that of HD 63433, and adopt the Sun's spectrum between 1216 \AA{} and 2588 \AA{}.

\subsection{Final reconstructed stellar spectrum}
\begin{figure}
    \includegraphics[width=0.5\textwidth]{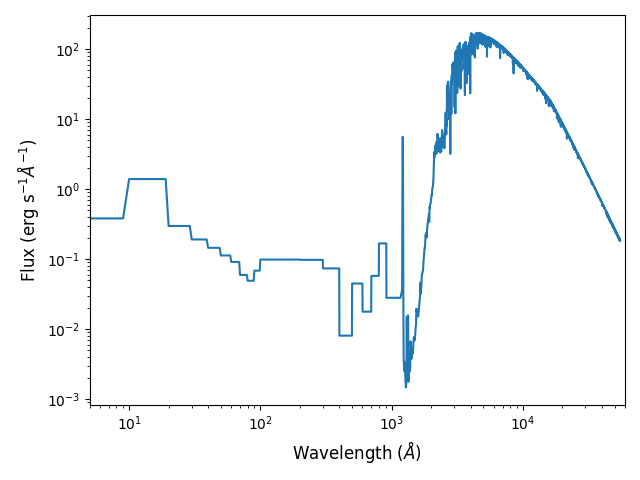}
    \caption{Fiducial stellar spectrum.  We binned the X-ray data for better visibility.}
    \label{fig:stellar_spectrum}
\end{figure}

To recap, we obtain the X-ray spectrum by fitting a model to XMM EPIC data; the EUV spectrum using the scaling relations of \cite{linsky_2014}; the Ly$\alpha$ spectrum by adopting ISM absorption parameters inferred from nearby stars and modifying a similar star's spectrum to fit the STIS data; the 1216--2588 \AA{} spectrum by assuming it is identical to solar; and the NUV, optical, and IR spectrum from PHOENIX.  The reconstructed spectrum is plotted in Figure \ref{fig:stellar_spectrum}.

\begin{table}[ht]
  \centering
  \caption{Band-integrated fluxes}
  \begin{tabular}{c C C}
  \hline
  	  Band & \text{Wavelengths (\AA)} & \text{Flux at 1 AU (cgs)}\\
      \hline
      X-ray & 5-100 & 27 \pm 14\\
      EUV & 100-912 & 91 \pm 27\\f
      Ly$\alpha$ & 1214-1217 & 56 \pm 17\\
      MUV & 1230-2588 & 2600 \pm 150\\
      Total & 5-50,000 & 1.02 \pm 0.04 \times 10^6\\
      \hline
  \end{tabular}
  \label{table:band_fluxes}
\end{table}

In Table \ref{table:band_fluxes}, we list the band-integrated fluxes of physical interest.  In addition to the nominal values, we make an attempt to estimate the error bars.  For the X-ray flux, we adopt 50\% errors because of the significant variability we see in even our 6 ks XMM observation.  The Ly$\alpha$ error is estimated based on \cite{linsky_2014}, who state that the Ly$\alpha$ reconstruction process gives rise to 10--30\% errors (we adopt 30\%).  The EUV error is also estimated based on \cite{linsky_2014}, who find that the scaling relations we relied on to obtain our EUV spectrum are accurate to 30--60\% (RMS) in 10 nm bins.  We assume that the errors bin down when the total EUV flux is calculated, and adopt 30\% as the final uncertainty.  The MUV error is calculated from the 7\% mismatch between the solar spectrum and the XMM OM photometry.  The nominal value and the error for the bolometric flux are both calculated using the luminosity from \cite{mann_2020}.

\section{Mass Loss Modeling}
\label{sec:modeling}
\subsection{Planet parameters}
\label{subsec:mass}

\begin{table}[ht]
  \centering
  \begin{tabular}{C C}
  \hline
  	  Parameter & Value\\
      \hline
      T_* & 5640 K\\
      R_* & 0.912 \pm 0.07 R_\Sun \\
      M_{p,c} & 7.3 M_\Earth \\
      M_{p,b} & 5.5 M_\Earth \\
      R_{p,c} & 2.67 R_\Earth \\
      R_{p,b} & 2.15 R_\Earth \\
      a_c & 0.1458 AU\\
      a_b & 0.0719 AU\\
      P_c & 20.5 d\\
      P_b & 7.1 d\\
      \hline
  \end{tabular}
  \caption{Stellar and planetary properties.  All values are from the discovery paper, \cite{mann_2020}.}
  \label{table:properties}
\end{table}

In order to set up our simulations of the outflow from each planet, we need to know the planets' radii, semimajor axes, masses and rotation periods.  These are summarized in Table \ref{table:properties}.  We obtain the first two from the discovery paper.  Unfortunately, the youth and high activity of HD 63433 make it difficult to measure planet masses using the radial velocity technique.  This is especially true for b, which has an orbital period close to the rotation period of the star.  The planets are also not particularly close to any orbital resonances, making it unlikely that we could obtain dynamical mass constraints using transit timing variations.  The assumed planet mass is a key ingredient for our models, as the predicted mass loss rate is exponentially sensitive to the assumed mass (e.g., Eq. 64 of \citealt{adams_2011}: $\dot{m} \propto \exp(-GM_p/R_p c_s^2)$). 

In the absence of any empirical mass constraints, we instead utilize a mass-radius relation derived from population-level studies of planets orbiting older (approximately greater than a Gyr) stars.  In the discovery paper for this system, \cite{mann_2020} used the \cite{chen_2016} probabilistic forecasting relation to calculate an estimated mass of $5.5 \pm 2 M_\Earth$ for b and $7.3 \pm 2 M_\Earth$ for c.  If we instead utilize the polynomial mass-radius relation from \cite{wolfgang_2016}, 
we would predict a mass of $7.3 \pm 2 M_\Earth$ for b and $9.7 \pm 2 M_\Earth$ for c.  If we use the mass-radius relation from \cite{bashi_2017}, we would predict a mass of 5.6 $M_\Earth$ for b and 8.2 $M_\Earth$ for c.  However, these mass-radius relations are all derived from observations of planets that are significantly older than HD 63433, whereas the planets in this system might still be inflated because they have lost a smaller fraction of their primordial atmospheres.  We therefore adopt the lowest mass estimates, those of \cite{chen_2016}, as the fiducial case for our models.  Using the scaling relation from the previous paragraph, we estimate that an uncertainty of 2 $M_\Earth$ in the estimated planet masses translates to an uncertainty of $\sim50\%$ in mass loss rate for c and $\sim80\%$ for b, assuming a sound speed of 10 km/s.

We next consider whether or not HD 63433 b/c are likely to be tidally synchronized.  The tidal synchronization timescale is \citep{guillot_1996}:

\begin{align}
    \tau_s = \frac{Q \omega_p M_p a^6}{G M_*^2 R_p^3}
\end{align}

Adopting a tidal quality factor of Q=100 and an initial rotation rate of one Earth day, this evaluates to 0.02 Myr for b and 0.7 Myr for c (i.e., much less than the present-day age of the star).  While Q is highly uncertain and the stronger gravity of a mini Neptune could reduce tidal dissipation by a factor of 2-3 \citep{efroimsky_2012}, it is difficult to get Q much above 1000 for a rocky planet \citep{clausen_2015}.  We therefore conclude that it is very likely that both planets have rotation periods equal to their orbital periods, and use this assumption in our models.

\subsection{Important physical processes}
\label{subsec:physical_processes}
Thermal mass loss is driven largely by stellar X-ray and EUV flux.  These high energy photons are absorbed far above the optical photosphere, heating the thermosphere.  The main cooling mechanism is emission of Ly$\alpha$ radiation by collisionally excited atoms \citep{murray-clay_2009}, which scales with temperature as $\Lambda \propto \exp(-118,348 K/T)$.  The strong exponential dependence of the cooling rate keeps the temperature around a few thousand K, but below $\sim10^4$ K.  In the region of the outflow closest to the planet, the temperature rises with increasing radius.  This is driven by the increased XUV heating and the decreased radiative cooling at larger separations.  Eventually, the importance of these effects diminishes as the optical depth of the atmosphere above becomes effectively transparent to XUV, and adiabatic cooling causes the temperature to slowly drop.  

Hydrodynamic outflows can be divided into three regimes, depending on the dominant mechanism for creating neutral hydrogen \citep{lampon_2021}.   In the outflow, stellar flux blueward of the Lyman limit destroys neutral hydrogen by ionizing it, while recombination creates it.  The neutral hydrogen population is also augmented by advection from lower in the atmosphere.  There are three possible regimes \citep{lampon_2021}, depending on whether recombination or advection is the dominant neutral hydrogen creation mechanism.  If recombination is dominant, the planet is in the recombination-limited regime, characterized by a narrow partially ionized zone and by low ($\sim5\%$) efficiency in converting flux to kinetic energy (because the energy is radiated away by recombination).  If advection is dominant, the planet is in the photon-limited regime, characterized by a very wide partially ionized zone and high flux-to-kinetic-energy conversion efficiency ($\sim15$\%).  If neither mechanism is clearly dominant, the planet is in the energy limited regime.  As we will discuss below, our simulations show that both HD 63433 planets are in the photon limited regime (Figure \ref{fig:rates}).  In this regime, the mass loss rate is high and neutral hydrogen is abundant in the outer regions, boosting the Ly$\alpha$ signal.

Helium absorption is from helium atoms in the metastable triplet ground state.  Helium must stay in this triplet ground state, 19.8 eV above the singlet ground state, in order to absorb at 1083 nm.  This level is populated, in most situations, by ionization followed by recombination, which ends with the recombining electron in the triplet state $\sim 3/4$ of the time.  Triplet helium can be destroyed by collisional de-excitation with electrons and neutral hydrogen in the lower atmosphere and by photoionization ($h\nu > 4.8~{\eV}$) in the upper atmosphere.  Other ways of producing and destroying triplet helium exist, such as collisional excitation and spontaneous radiative decay, but they are usually negligible.  Close to the planet, no ionizing radiation penetrates, so nothing is ionized and there is no recombination to populate the triplet state.  Going outwards, the electron density at first increases, causing the triplet state number density to increase to some maximum.  Past this maximum, where the atmosphere is already largely ionized, $n(e)$ and $n({\rm He II})$ both decrease with $r$ because of expansion.  In this region, collisional de-excitation falls with $r$, but not enough to overcome the decreased production and the mostly constant photoionization of triplet helium, causing the triplet number density to fall.  For a more detailed, quantitative overview of the physics of helium absorption in exoplanets, see \cite{oklopcic_2018}.

Although we have discussed these processes as if they are radially symmetric, in reality outflows will have a non-spherical geometry that is shaped by the planet's immediate environment.  The stellar wind and radiation pressure both work to push escaping gas away from the star.  The Coriolis force then imparts a sideways force to the gas, creating a comet-like tail trailing the planet \citep{schneiter_2007}.  If the planet has a magnetic field it can suppress the mass loss rate and confine the outflow (e.g. \citealt{adams_2011,owen_2014,khodachenko_2015}), but exoplanetary magnetic fields are poorly understood because they have never been observed.  We simulate the outflow from both planets without including magnetic fields, and discuss the ways that magnetic fields might alter our predictions in \S\ref{subsec:magnetic_fields}.

\subsection{3D hydrodynamic models}
Our 3D models utilize the approach outlined in \citet{wang_2018,wang_2020}, which combines ray-tracing radiative transfer, real-time non-equilibrium thermochemistry, and hydrodynamics based on the higher-order Godunov method code \verb|Athena++| \citep{stone_2020}.  We include a stellar wind with a mass-loss rate of $\sim 8\times$ solar, as calculated in \S\ref{subsec:magnetic_fields}, and a roughly solar velocity of $400~{\rm km\ s}^{-1}$.  Our models account for the hydrodynamic and thermochemical interactions between the stellar wind and the planetary outflow.  These 3D models capture the anisotropy of the outflow pattern better than the 1D LTE models, while the non-LTE thermochemistry self-consistently predicts the mass loss rate and the line profiles. The model incorporates a total of 26 species and 135 reactions, including various relevant heating and cooling processes (e.g. photoionization, photodissociation of molecular hydrogen,  Lyman-$\alpha$ cooling, etc.).  Metastable helium is a chemical species like the others, and key reactions that form and destroy this species are included in the thermochemical network.  We refer the reader to \cite{wang_2018} and \cite{wang_2017} for details of these reactions.  The most important cooling processes in the models presented here are recombination, PdV work, and ro-vibrational cooling by H$_2$O/OH and CO, while the most important heating mechanism is photoionization of H and He (see Figure 3 in \citealt{wang_2018}).

Starting from the stellar spectrum computed in Section \ref{sec:understanding_star}, we group photons into seven energy bins for the ray-tracing calculation:
\begin{itemize}
    \item $h\nu = 1.4~\eV$ for infrared, optical, and near ultraviolet (NUV) photons, $h\nu = 7~\eV$ for ``soft'' far ultraviolet (FUV) photons
    \item $h\nu = 12~\eV$ for the Lyman-Werner band FUV photons, which can photodissociate molecular hydrogen but cannot ionize them (this is not to be confused with the Ly$\alpha$ line, which the band does not include)
    \item $h\nu = 16~\eV$ for ``soft'' extreme ultraviolet (soft EUV) photons, which can ionize hydrogen but {\it not} helium
    \item $h\nu = 47~\eV$ for hard EUV photons that ionize hydrogen {\it and} helium
    \item $h\nu = 300~\eV$ for soft X-rays, which are abundant for an active star like HD 63433
    \item $h\nu = 3000~\eV$ for hard X-rays
\end{itemize}

We note that this discretization of radiation artificially shrinks the vertical extent of the region where significant stellar XUV is deposited.  If the radiation were not discretized, the photoionization cross section would drop with photon energy beyond the ionization energy, so that higher energies take over as the outflow becomes optically thick to lower energies.  Unfortunately, this discretization is required in order to make our 3D model computationally tractable.

In addition to the opacities caused by photochemical reactions, we also include an effective opacity term in all bands, using dust/PAH as a proxy \citep[see also][]{wang_2018}.  This is particularly important in the optical band due to its heating effects, as our opacity calculation did not include the Thomson cross-section $\sigma/\chem{H}\simeq 6.7\times 10^{-25}~\cm^2$.

\begin{deluxetable}{lr}
  \tablecolumns{2} 
  \tabletypesize{\normalsize}
  \tablewidth{0.5\textwidth}
  \tablecaption
  { Properties of the fiducial model}   
  \tablehead{
    \colhead{Parameter} &
    \colhead{Value}
  }
  \startdata
  \emph{Planet Interior} & \\
    $R_{\rm core,c}$ & $2.62~R_\Earth$ \\
    $R_{\rm core,b}$ & $1.958~R_\Earth$ \\
  \\  
   \emph{Mass loss} & \\
     $\dot{M}_{\rm b}$
     & $0.35~M_\Earth~{\rm Gyr}^{-1}$ \\
     $\dot{M}_{\rm b, 10\ M_\oplus}$
     & $0.17~M_\Earth~{\rm Gyr}^{-1}$ \\
     $\dot{M}_{\rm c}$
     & $0.11~M_\Earth~{\rm Gyr}^{-1}$ \\
   \\  
  \emph{Simulation Domain} & \\
  Radial range (c) & $0.98 \le (r/R_p) \le \ 94$\\
  Radial range (b) & $0.91 \le (r/R_p) \le \ 116$\\
  Latitudinal range & $0\le\theta\le\pi$ \\
  Azimuthal range & $0\le\phi\le\pi$ \\
  Resolution $(N_{\log r}\times N_{\theta} \times
  N_{\phi})$ & 
  $192\times 128\times 64 $ \\ 
  \\
 \emph{Photon Flux at 1 AU$^*$} [$\cm~^{-2}~\s^{-1}$] & \\
  $1.4~\eV$ (IR/optical)  & $4.5\times 10^{17}$ \\
  $7~\eV$ (Soft FUV)   & $1.8\times 10^{16}$ \\
  $12~\eV$ (LW)   & $3.0\times 10^{11}$ \\
  $16~\eV$ (Soft EUV) & $1.2\times 10^{12}$ \\
  $47~\eV$ (Hard EUV)  & $3.2\times 10^{11}$ \\
  $0.3~\keV$ (Soft X-ray) & $4.7\times 10^{10}$ \\  
  $3~\keV$ (Hard X-ray)  & $1.5\times 10^{9}$ \\  
  \\
  \emph{Initial Abundances} [$n_{\chem{X}}/n_{\chem{H}}$] & \\
  \chem{H_2} & 0.5\\
  He & 0.1\\
  \chem{H_2O} & $1.8 \times 10^{-4}$\\
  CO & $1.4 \times 10^{-4}$\\
  S  & $2.8 \times 10^{-5}$\\
  Si & $1.7 \times 10^{-6}$\\
  Dust grains & $1.0 \times 10^{-9}$ \\
  \\
  \emph{Dust/PAH Properties} & \\
  $\sigma_\dust/\chem{H}$  & $8\times 10^{-22}~\cm^2$\\
  $m_{\rm dust} / m_{\rm gas}$ & 7\e{-7}\\
  \\
  \emph{Stellar Wind} & \\
  $\dot{M}$ & 2\e{13} g/s\\
  $v_{\rm w}$ & 400 km/s\\
  Temperature & $10^6$ K\\
  \enddata
  \tablecomments{$^*$Divide by 0.1458$^2$ for c and 0.0719$^2$ for b}
  \label{table:simulation-fiducial}
\end{deluxetable}

A typical planetary atmosphere consists of a convective interior and a quasi-isothermal exterior \citep[e.g.][]{rafikov_2006}.  With the use of an interior model, we adjusted the envelope fraction to match the observed radius and assumed mass of the planets. We find an envelope mass fraction of $\sim0.6\%$ for b and 2\% for c, typical values for mini Neptunes.  However, we note that without a precise mass measurement, the mass fraction of the H/He envelope cannot be precisely constrained. The precise envelope fraction is not important to modelling the outflow because it is the gravity and density at the optical transit radius that set the inner radial boundary condition, above which all photospheres of relevant bands of radiation are located. We summarize the key quantities that define the fiducial model in Table~\ref{table:simulation-fiducial}. Note that, to reduce the cost of simulation while keeping all necessary physical features, we only simulate the half-space above the orbital plane, and assume reflection symmetry over that plane.

\subsection{Model results}
\label{subsec:model_results}
\begin{figure*}
  \centering 
    \subfigure {\includegraphics
    [width=0.33\textwidth]{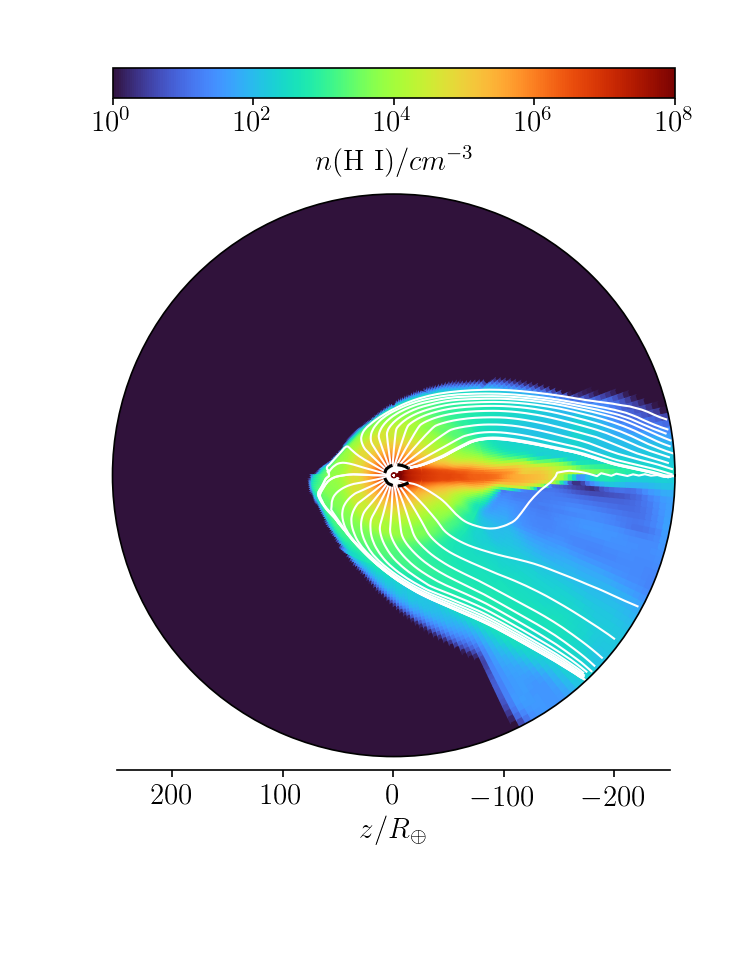}}\subfigure {\includegraphics
    [width=0.33\textwidth]{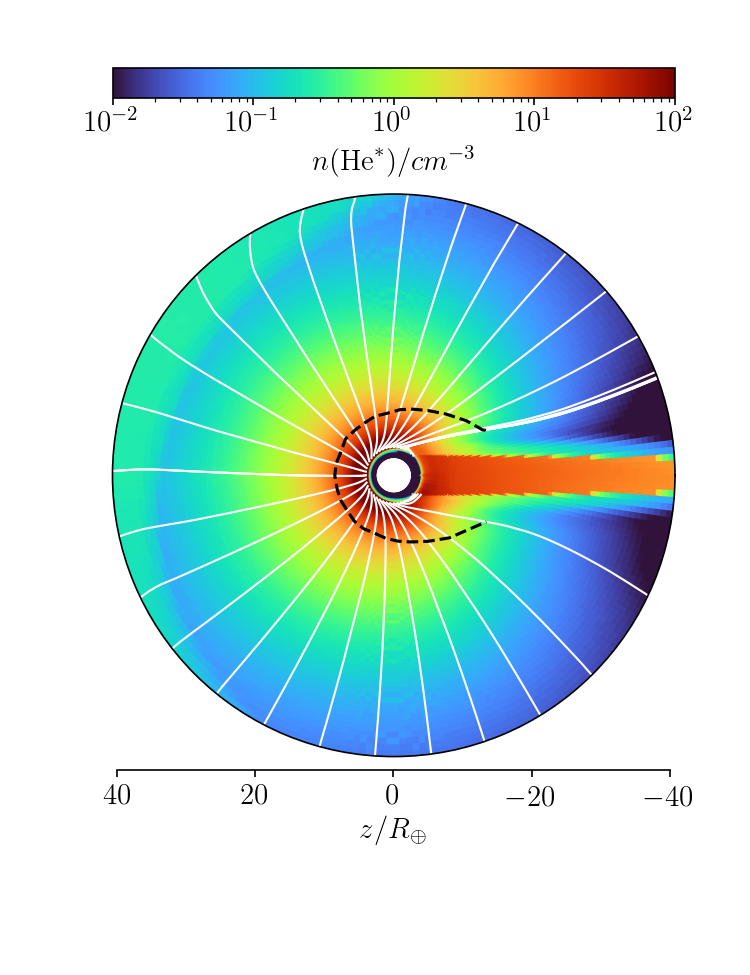}}\subfigure {\includegraphics
    [width=0.33\textwidth]{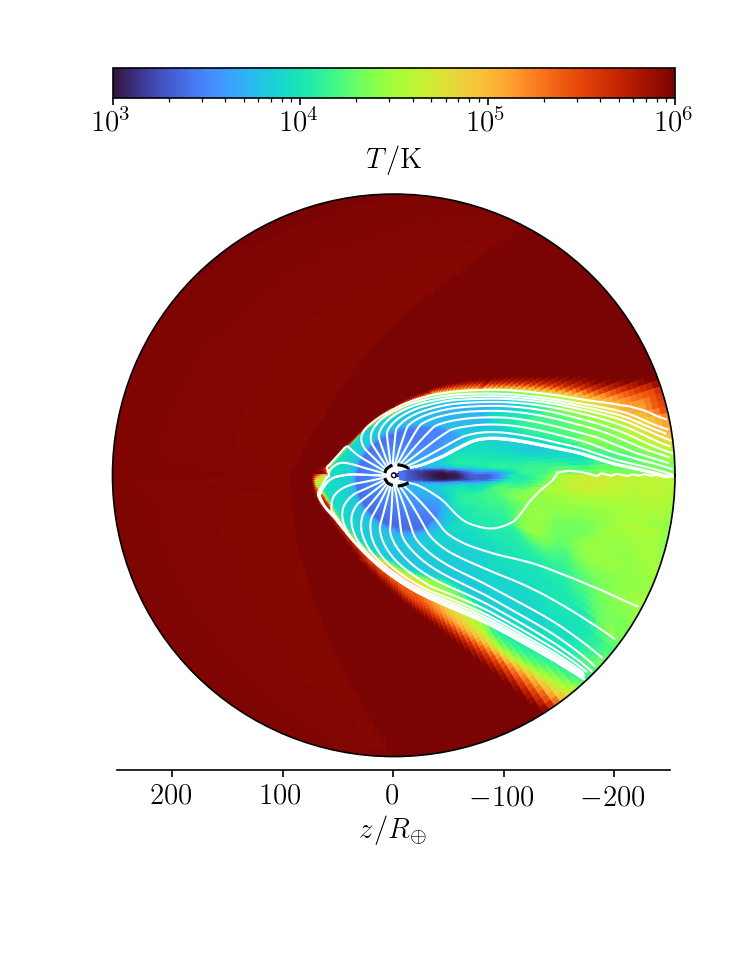}}
    \subfigure {\includegraphics
    [width=0.33\textwidth]{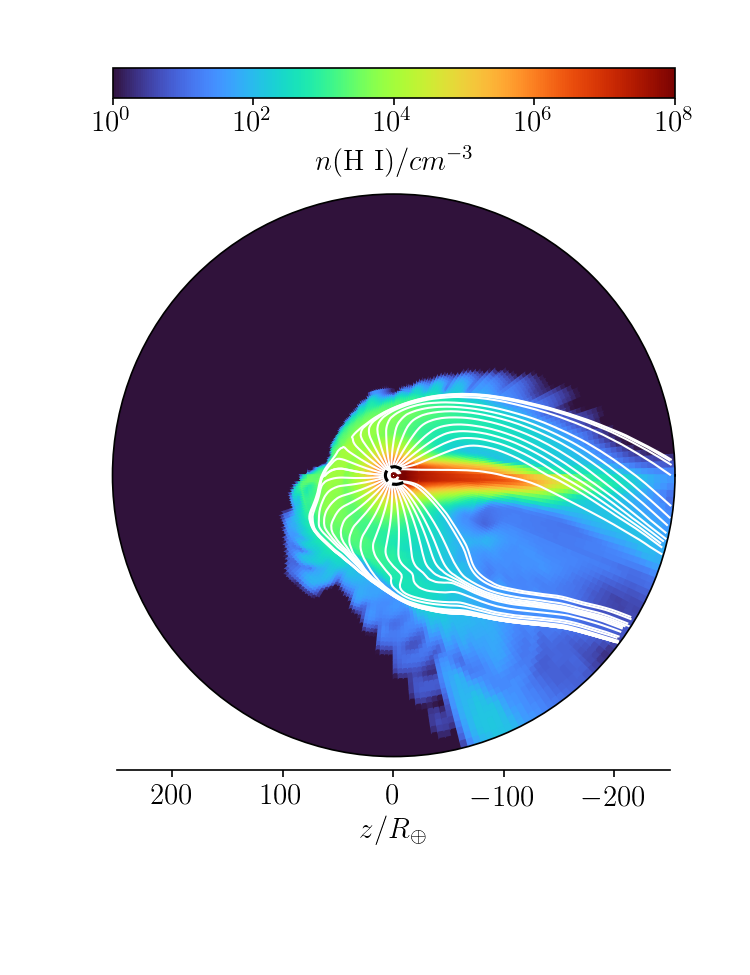}}\subfigure {\includegraphics
    [width=0.33\textwidth]{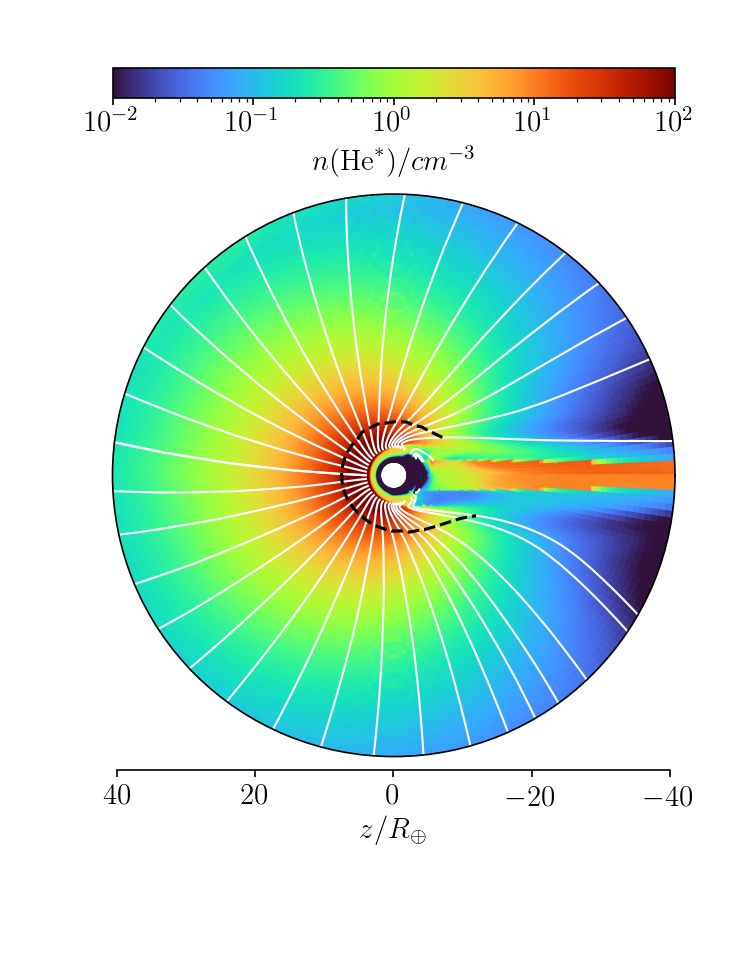}}\subfigure {\includegraphics
    [width=0.33\textwidth]{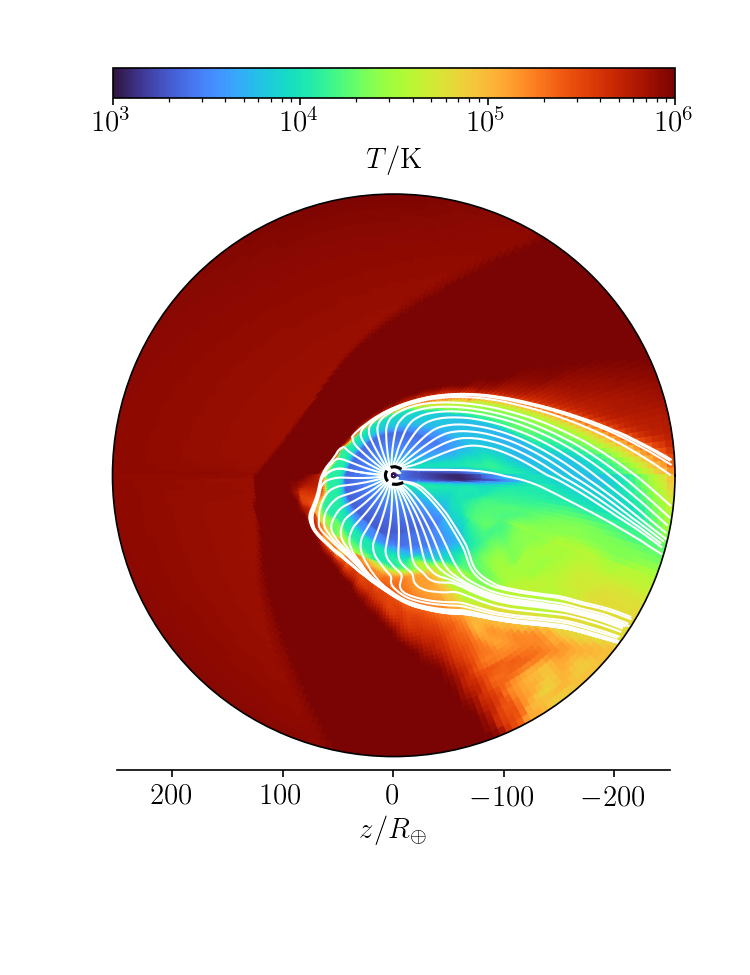}}
    \caption{Neutral hydrogen number density, temperature, and triplet helium density from the time-averaged (over the last $\sim 10$ kinematic timescales) fiducial 3D models, for c (top) and b (bottom).  The star is toward the left, the velocity vector of orbital motion lies in the paper plane and points upwards. These plots show the profiles in the orbital plane. The white lines are the streamlines, while the dashed black lines represent the inner sonic surface. Note that the middle column showing metastable helium distribution zooms into the innermost $40~R_\oplus$ region for clearer presentation. }
    \label{fig:slices}
\end{figure*}

\begin{figure}
  \centering 
  \subfigure {\includegraphics
    [width=0.47\textwidth]{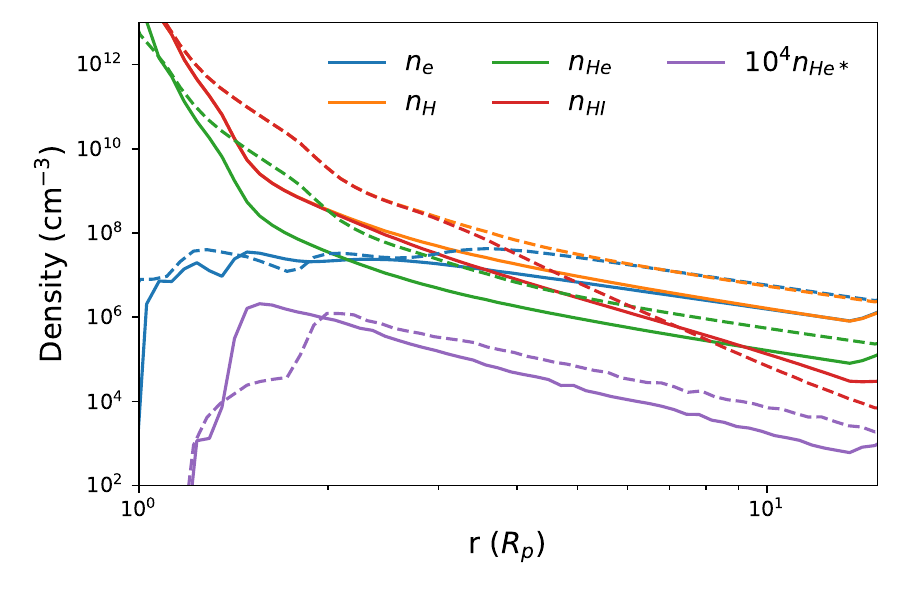}}
    \subfigure {\includegraphics
    [width=0.47\textwidth]{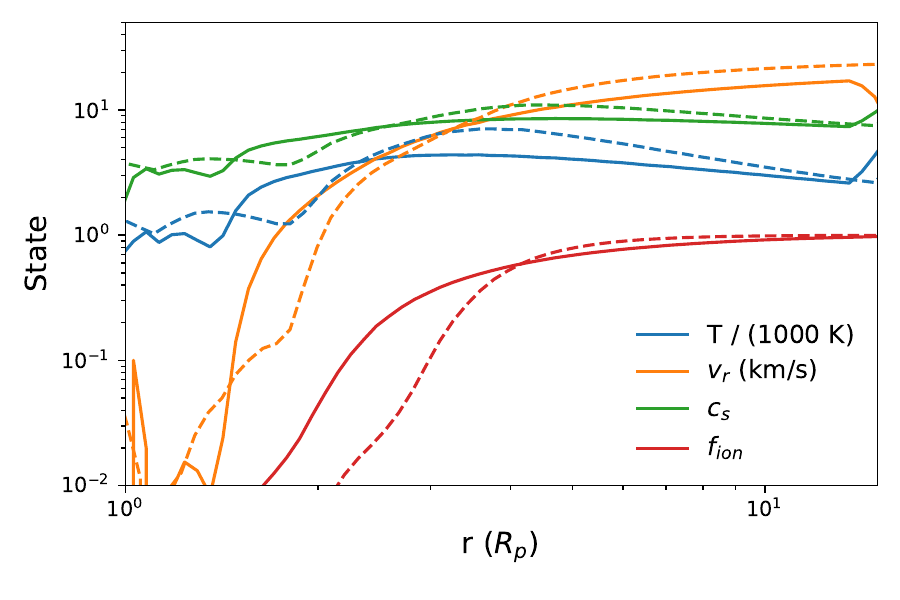}}
    \caption{Radial profiles of key quantities along the radial lines pointing to the direction of orbital motion. Profiles are calculated for the time-averaged data (over the last $10$ kinematic timescales of the simulations) for the fiducial models of planets c and b. Top: number densities of electrons ($n_e$), neutral helium ($n_{\rm He}$), metatstable helium ($n_{{\rm He}^*}$), neutral hydrogen ($n_{\rm H}$), and ionized hydrogen ($n_{HI}$).  Bottom: temperature $T$, radial velocity $v_r$, sound speed $c_s$, and hydrogen ionization fraction $f_{\rm ion}$.  Solid lines are for planet c; dashed lines for b.}
\label{fig:densities_states}
\end{figure}

\begin{figure}
  \centering 
  \subfigure {\includegraphics
    [width=0.47\textwidth]{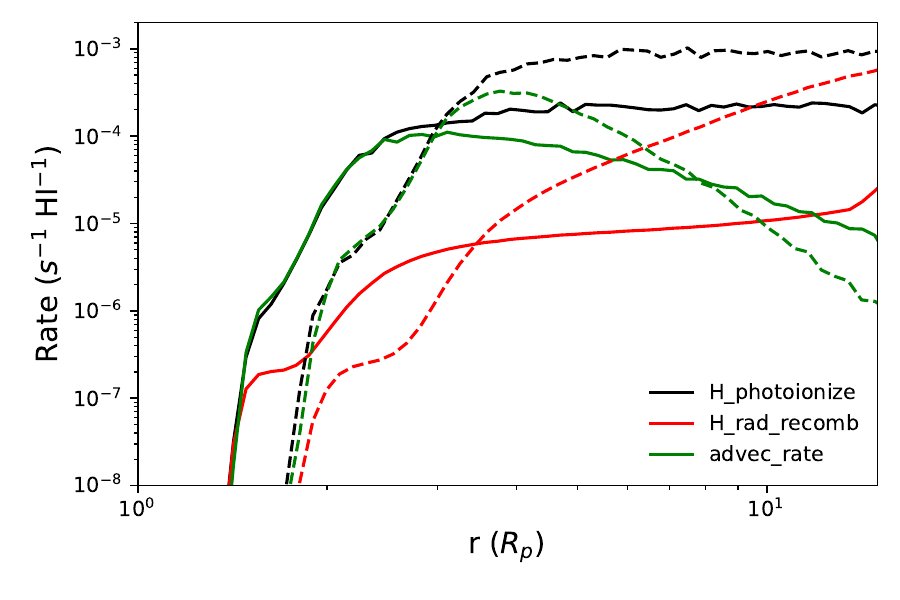}}
    \subfigure {\includegraphics
    [width=0.47\textwidth]{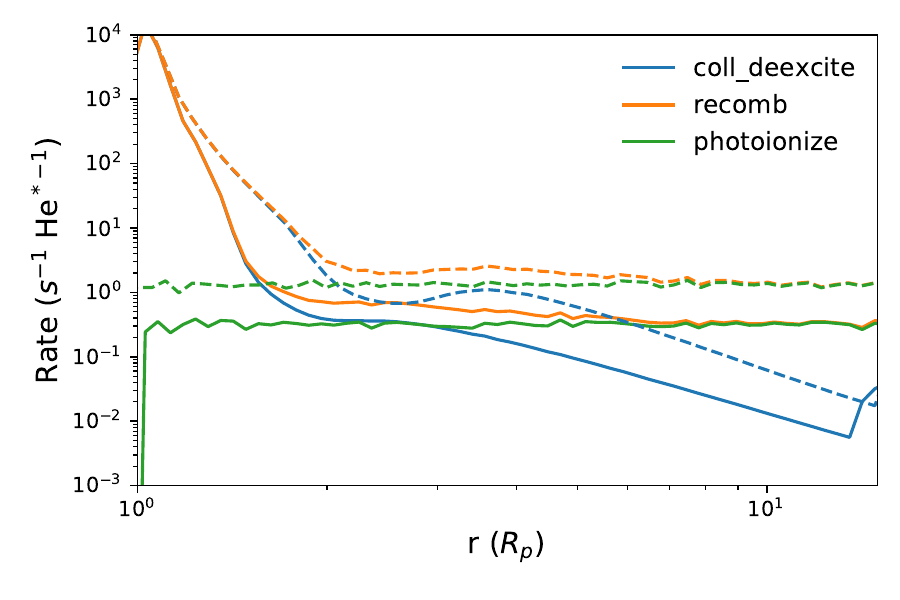}}
    \caption{Similar to Figure~\ref{fig:densities_states} but for important reaction rates. Top: rates of important processes that create or destroy neutral hydrogen, namely photoionization, radiative recombination, and advection.  Bottom: rates of processes that create or destroy metastable helium, namely collisional de-excitation, recombination, and photoionization [${\rm He}^* + h\nu(>4.8~{\rm eV}) \rightarrow {\rm He} + e^-$]. Rates for planet c are shown in solid lines, and those for planet b are in dashed lines.  The rates are computed along the radial line pointing to the direction of the planets' orbital motion.}
\label{fig:rates}
\end{figure}

Our models predict a mass loss rate of 0.11 $M_\Earth$/Gyr (2.1\e{10} g/s) for c and 0.35 $M_\Earth$/Gyr (6.6\e{10} g/s) for b. We can use our initial envelope fractions (2\% for c and 0.6\% for b) to calculate corresponding atmospheric mass loss timescales of 0.9 Gyr for planet c and 0.08 Gyr for planet b.

Figure \ref{fig:slices} shows the orbital plane of our simulations.  The outflow is initially somewhat spherically symmetric and still maintains this symmetry at $\sim10R_\Earth$, the approximate photospheric radius for metastable helium absorption.  Around 50 $R_\Earth$ it loses this symmetry as gas emanating from the day side is pushed toward the night side by the stellar wind and radiation pressure.  As we move radially outward from the planet in a direction perpendicular to the planet-star axis, the temperature is initially equal to the planetary equilibrium temperature, rises to a peak of a few thousand Kelvin, declines slightly, and then jumps to a few million Kelvin as the outflow encounters the 1 MK stellar wind in a shock.

Figure \ref{fig:densities_states} (top) shows the number densities of various species radially outward from the planet along the direction of orbital motion. The bottom pane shows the temperature, radial velocity, sound speed, and hydrogen ionization fraction.  We find that the triplet helium density, which controls helium absorption, peaks around 200 cm$^{-3}$ at 1.6 $R_p$ (c) or 120 ${\rm cm}^{-3}$ at 2.1 $R_p$ (b) before slowly declining farther out.  The neutral hydrogen density declines smoothly with increasing distance.  The temperature rises to a maximum of 4000 K at 3 $R_p$ (c) or 7000 K at 4 $R_p$ (b) before slowly declining.  The sound speed hovers around the 7-10 km/s typical of ionized hydrogen at several thousand Kelvin.  At larger radii ($>5 R_p$), the outflow velocity asymptotes to $\sim$2 times the sound speed and the density falls roughly as $r^{-2}$, in accordance with the analytic Parker wind prediction.  The neutral hydrogen fraction is nearly 1 at the surface, but declines to 50\% around 3 $R_p$, falling to $<$3\% at 15 $R_p$.

The rates we plot in Figure \ref{fig:rates} show that triplet helium is created by recombination and destroyed predominantly by collisional de-excitation (smaller radii) or photoionization (larger radii), as discussed in \S\ref{subsec:physical_processes}.  This is the same qualitative behavior seen for 55 Cnc e with a 1D PLUTO-CLOUDY model \citep{zhang_2020}, for some generic planets with a Parker wind model \citep{oklopcic_2019}, and for the gas giant WASP-107b with the same 3D model \citep{wang_2020}. Our models do include collisional excitation and radiative decay, but they are not plotted because they are negligible.  Collisional excitation is negligible because the collisional excitation coefficient is 12 orders of magnitude smaller than the collisional deexcitation coefficient \citep{oklopcic_2018}.  Although singlet helium is typically $\sim 10^6$ times more abundant than triplet helium, a gap of $10^6$ remains.  Radiative decay is negligible because transitions between the triplet and singlet states are forbidden.  This means that the Einstein $A$ coefficient for the transition to the singlet ground state is very small (1.272\e{-4} s$^{-1}$) and the decay timescale is very long ($\sim 2.2~{\rm hr}$).  We can compare this decay timescale to the triplet helium production timescale, which is the triplet number density divided by the recombination rate.  This timescale is on the order of $0.5~{\rm s}$.  This short production timescale means that the triplet helium density is in local equilibrium, and is not significantly affected by advection except indirectly (advection carries neutral gas outward, reducing the electron number density).

Figure \ref{fig:rates} shows the rates of various processes that create and destroy neutral hydrogen: photoionization, recombination, and advection.  It can be seen that for both planets, advection dominates over recombination until the outermost regions of the outflow (11 $R_p$ for c, 7 $R_p$ for b). We conclude that both planets are in the energy limited regime for photoevaporative mass loss, where photoevaporation is relatively efficient because recombination does not radiate away all the incoming stellar high-energy radiation \citep{lampon_2021}.  The rate for photoionization plateaus around 2.5 (3.5) $R_p$ for planet c (b) because the outflow becomes optically thin to ionizing radiation.  This is the same general region where the flow becomes supersonic and where the hydrogen becomes predominantly ionized, making recombination and collisional de-excitation with electrons increasingly important.   The triplet helium fraction, however, peaks well before this critical point (1.6 $R_p$ for c and 2.1 $R_p$ for b) before beginning a decline of $n_{He*} \propto r^{-3.5}$ for c and $n_{He*} \propto r^{-3}$ for b.  

Our models predict a relatively symmetric transit light curve in both Ly$\alpha$ (Figure \ref{fig:lyman_alpha_lc}) and helium (Figure \ref{fig:lines}).  This stands in contrast to Ly$\alpha$ observations of GJ 436b \citep{lavie_2017} and helium observations of WASP-107b \citep{allart_2019,kirk_2020}, which show a much more delayed and extended egress.  This happens because the stellar wind and radiation pressure both push the outflow away from the star, where the Coriolis acceleration $-2\mathbf{\Omega} \times \mathbf{v}$ slows its velocity relative to the planet's orbital motion.  Naively, one would expect a stronger stellar wind to cause a longer tail and more asymmetric transit shape, but this is not the case in our models.  We initially performed 3D simulations with a solar-strength wind and saw an asymmetrical helium transit, with peak absorption occurring 1.5 hours after the white light transit mid-point for planet c.  When we switched to a more realistic 8$\times$ solar wind, we obtained the fiducial model presented here.  In this version of the model the confining effect of the stellar wind overpowers the Coriolis force, accelerating the outflow and increasing the importance of inertial forces relative to Coriolis forces (parameterized by the Rossby number, Ro=vr/(2$\Omega$)).  This results in a more symmetric transit shape for the 8$\times$ solar wind case.

\section{Discussion}
\label{sec:discussion}
\subsection{Comparing model and data}
Before comparing our model predictions to the observed magnitude of absorption during transit, it is useful to consider the implications of the atmospheric mass loss timescales in these models.  Although these quantities are sensitive to the assumed mass, we can nonetheless draw some general conclusions.  Planet c has a mass loss timescale of 0.9 Gyr in our models, while b has a mass loss timescale of 0.08 Gyr.  The star has an estimated age of 0.4 Gyr, suggesting that b is unlikely to have retained a primordial atmosphere while it is at least plausible for c to have done so.  At earlier times, the planets' puffier radii leads to lower gravity and a higher Roche radius, both leading to increased mass loss. This prediction is consistent with our non-detection of excess absorption in either Ly$\alpha$ or metastable helium absorption during transits of planet b.  We explore the effect that the assumed mass of planet b has on its predicted atmospheric lifetime in more detail in \S\ref{subsec:mass_of_planet_b}.

If we set these arguments aside for the moment and assume that both planets host hydrogen-rich atmospheres, we can use our 3D models to predict the time-dependent absorption signal during transit.  Figure \ref{fig:lyman_alpha_lc} shows our model predictions for the observed Ly$\alpha$ absorption.  We find that the model slightly underpredicts the amount of blue wing absorption for planet c.  The model also predicts that the point of maximum absorption will occur slightly after mid-transit, whereas our data prefers a peak before mid-transit.  The model predicts that there should be negligible absorption in the red wing bandpass; although we see weak evidence for red wing absorption in our data, the detection is not conclusive (see \S\ref{subsec:ly_alpha_analysis} for more details).  For planet b, the model predicts an absorption depth in the blue wing of the Ly$\alpha$ line that is comparable in magnitude to that of planet c, with a shorter but symmetric transit shape.  This absorption signal is conclusively ruled out by our data.  As with c, the model predicts negligible absorption in the red wing; this is consistent with our non-detection.

For planet c, where we detect absorption in the blue wing of Ly$\alpha$, we can compare the predicted wavelength-dependent shape of the absorption signal as a function of time to the observed line shape (Fig. \ref{fig:excess_spectrum_c}).  The magnitude of absorption in the observed spectrum rises from $<$5\% to 25\% as we move from -100 km/s to -10 km/s, while the model spectrum rises from $<$5\% to 60\% as we move from -50 km/s to -10 km/s.  We conclude that the observed absorption signal is significantly more widely dispersed in velocity space than the modeled absorption.  This phenomenon of Ly$\alpha$ absorption being present far above the sound speed was first noted by \cite{vidal-madjar_2003} for HD 209458b, and has also been observed in all four planets with definitive Ly$\alpha$ detections (HD 209458b, HD 189733b, GJ 436b, GJ 3470b).  \cite{holmstrom_2008} proposed that the -140 km/s absorption signal seen for HD 209458b is caused by charge exchange between neutral hydrogen in the planetary outflow and solar wind protons, creating a population of neutral hydrogen atoms with a high velocity dispersion heading away from the star. We note that kinetic processes in the interaction region between the stellar wind and the planetary outflow might be difficult to model correctly due to the long mean free paths, which exceed 1 $R_p$ when $n_H <\sim 10^4$ cm$^{-3}$. If future models deal with this process with more consistency, the observed signal might be properly explained by invoking stellar wind hydrogen ions neutralized and thermalized near the stellar wind-planetary outflow contact surface.

While the simulation explains the blue wing absorption of c somewhat well, it does not explain the red wing absorption at all.  The simplest explanation for the discrepancy is that the observed red wing absorption is actually stellar variability.  The absorption detection is marginal even from a pure SNR perspective, and there are other reasons to doubt the detection (Subsection \ref{subsec:ly_alpha_analysis}), including the non-detection in the alternative analyses (see Appendix).  If we assume the detection is real, the red wing absorption is difficult to explain purely hydrodynamically because it occurs at 100--200 km/s, far above the sound speed of $\sim$10 km/s.  Since some extreme red wing absorption can be obtained from energetic neutral atoms at the intersection of the planetary outflow and the stellar wind, one might think the problem can be solved by making the stellar wind faster (resulting in more kinetic energy being thermalized in the shock) and denser.  However, this would also tend to confine the outflow and push more of the absorption to the blue wing, potentially worsening the discrepancy between model and data.  Using an order-of-magnitude calculation, we also considered the possibility that the red wing absorption could be explained by the far wings of Lyman alpha.  We considered a r=25 $R_\Earth$ uniform sphere with n(HI) = $10^6$ cm$^{-3}$ with T=7000 K and a mean atomic weight of 0.7 AMU, very roughly matching the simulated conditions in Figure \ref{fig:slices}.  We found that the optical depth across the center of the sphere is 0.03 at 50 km/s, 0.006 at 100 km/s, and 0.003 at 150 km/s.  Even an optical depth of 0.03 across the entire 25 $R_\Earth$ disk would increase the transit depth by only 0.2\%.  We conclude that if the absorption signal at 100-200 km/s is real, it is unlikely to be caused by gas moving at less than 50 km/s.

We next consider whether or not the predicted metastable helium absorption spectrum from these models is consistent with the upper limits from our NIRSPEC observations (see Fig. \ref{fig:helium_excess_planetary}).  For c, the model predicts 0.4\% absorption, which would be reduced to 0.15\% due to the self-subtraction when we remove the time-varying telluric signal.  This absorption should be marginally detectable, but may plausibly be hidden by stellar activity.  For b, the model predicts 0.5\% absorption, which would be reduced to 0.3\% by self-subtraction.  This is slightly stronger than the 0.2\% absorption in Figure \ref{fig:helium_excess_planetary}, which we attribute to stellar variability.  Our injection/recovery tests show that the predicted signal would have been detectable in both the residuals images and the excess absorption spectrum.  In addition to transit-averaged excess absorption spectra, we also compare the line-integrated light curves to model predictions (Figure \ref{fig:lines}).  Here, the model predicts a level of absorption considerably above the noise for both planets, but not considerably above the stellar variability we observe on both nights. 

\subsubsection{A higher mass cannot explain the non-detections for planet b}

We next consider whether or not a higher assumed mass for planet b can explain the Ly$\alpha$ and helium non-detections.  We run another 3D model with a core mass of 10$~M_\Earth$; this is 2.7$~M_\Earth$ higher than the predictions of all mass-radius relations we consider in \S\ref{subsec:mass}, and high enough for runaway accretion during planet formation to occur.  Aside from the increased core mass and correspondingly increased envelope fraction, all other aspects of the model were unchanged.

As expected, we find that increasing the assumed core mass does modestly reduce the magnitude of the predicted absorption signal during transit.  The high mass model predicts a Ly$\alpha$ absorption depth in the blue wing of 7\%, compared to 10\% in our fiducial model.  However, this signal is still too strong to accord with our observations.  In the helium lines, the high mass model predicts a peak absorption of 0.35\%, compared to 0.6\% in the fiducial model.  If we interpret the darkening of the helium line during the transit of planet b as planetary absorption instead of stellar variability, the observed signal would be comparable in magnitude to this prediction.  If the helium signal is planetary, it should be detectable with follow-up observations from telescopes even smaller than Keck, as we were limited by stellar variability and not by photon noise.

\subsubsection{Possible envelope compositions for planet b}
\label{subsec:mass_of_planet_b}

As previously noted, planet b's short predicted mass-loss timescale and our non-detection of absorption from either hydrogen or helium suggests that this planet may have already lost its primordial hydrogen/helium envelope.  Assuming an Earth-like composition, a bare core with the radius of b would have a mass of $M/M_\Earth = (R/R_\Earth)^4 = 21 M_\Earth$ \citep{lopez_2014}.  This core would be comparable in mass to the inferred cores of gas giant planets, and would have been highly susceptible to runaway accretion even at its present-day location in the inner disk \citep[e.g.,][]{lee_2019}.  Rocky cores with radii between 2-3 $R_\Earth$ planets are quite rare among the sample of planets with measured densities, although not unheard of \citep{mocquet_2014}.  This suggests that a small fraction of 10-40 $M_\Earth$ cores may avoid runaway gas accretion, presumably because they did not form until the disk was already dispersing \citep{lee_2019}.

If we are willing to consider high mean molecular weight envelopes, it is also possible to match planet b's observed radius with lower core masses.  Mass-radius relations for water-rich planets suggest that they can have bulk densities similar to those of mini Neptunes with rocky cores and hydrogen envelopes with mass fractions of $1-2\%$ (e.g. \citealt{turbet_2020,aguichine_2021,nixon_2021}).  Outflows from planets with water-rich envelopes will contain hydrogen created by the photodissociation of water, and hence will still absorb in Ly$\alpha$. \cite{johnstone_2020} found that the energy efficiency of photoevaporation for water atmospheres ($\sim$10\%) is similar to that of hydrogen/helium atmospheres.  However, up to 7/8 of the outflowing mass would be oxygen atoms, which would reduce the hydrogen number density and the corresponding magnitude of the absorption signal in Ly$\alpha$.  Water worlds would not have significant amounts of helium in their atmospheres, consistent with our upper limit on helium absorption.  More modelling is required to determine if our Ly$\alpha$ non-detection for b (or our detection for c) is consistent with a water world.  We might also consider high mean molecular weight envelopes with other compositions, such as CO$_2$ or N$_2$, although these atmospheres would be more compact and would require correspondingly larger core masses.  \cite{kite_2020b} found that highly irradiated planets are likely to lose heavy gases alongside hydrogen and helium, but that a volcanically revived atmosphere is plausible for $T_{eq} <\sim$ 1000 K planets around solar-mass stars.  These compact atmospheres might cool efficiently, which would suppress the outflow and reduce the observational signature of Ly$\alpha$, and there should be no helium because helium is insoluble in lava.  Such a metal-rich outflow might still be detectable in other FUV lines \citep{garcia_2021}.

\subsection{Other potentially important physical effects}

Our models of hydrogen-rich outflows are ultimately unable to provide a satisfactory match to the observational data for either planet.  While we can argue that planet b likely lost its primordial atmosphere, that is not the case for planet c.  We therefore consider whether or not other physical effects, such as stellar or planetary magnetic fields, that we neglect in our models might have a significant effect on the magnitude or shape of the predicted outflows.

\subsubsection{Magnetic fields}
\label{subsec:magnetic_fields}
In order to evaluate the potential importance of magnetic fields, we first need to estimate the stellar magnetic field strength.  \cite{vidotto_2014} measure the magnetic field strength of 73 stars ranging from 1 Myr to 10 Gyr and found that it is proportional to $t^{-0.655 \pm 0.045}$.  We use this scaling relation to estimate the magnetic field of the 440 Myr HD 63433.  We find that this star is predicted to have a magnetic field strength of approximately 5 times solar, or $\sim$13 G.  This is consistent with \cite{rosen_2016}, who used polarization data to measure the mean magnetic fields of five young Sun-like stars between 300 and 700 Myr and found $B_* \approx 20$G (with a range of 10-25 G), with no evidence for any age dependence.  We adopt 20 G as the fiducial field strength and assume that, in the absence of a stellar wind, its strength falls off as $r^{-3}$ (consistent with a dipole).  It should be noted that the mean field of the Sun varies from 0.2 to 2 G across a solar cycle \citep{plachinda_2011}, and it is likely that HD 63433 also has a variable field strength.  

If we assume the Elasser number $\Lambda = \frac{\sigma B^2}{\rho \Omega}$ is 1, where $\rho$ is the density of the interior, $\Omega$ is the rotation rate, and $\sigma$ is the conductivity, we can estimate the planetary magnetic fields.  Assuming that the interiors of the HD 63433 planets have Earth-like densities and conductivities, the magnetic field should scale as the square root of the rotation rate.   Rescaling Earth's magnetic field strength (0.25-0.65 G) to account for the slower rotation rates of the HD 63433 planets, we find that the planetary magnetic field is around 0.15 G for b and 0.08 G for c.  We further assume that the planetary magnetic field is a dipole and falls off with distance as $r^{-3}$.  We note, however, that the predicted magnetic field strengths of mini Neptunes are highly uncertain.  For example, \cite{christensen_2006} numerically analyze dynamo models and conclude that the Elasser number can range from 0.06 to 100 while the magnetic field is independent of rotation rate. \cite{christensen_2009} find using observations that B scales with $q_0^{2/3}$, where $q_0$ is the heat flux.  If so, the magnetic field of the HD 63433 planets could be greater than that of Earth.

Having estimated the relevant magnetic field strengths, we explore their importance for the outflow.  Following \cite{owen_2014}, we calculate the ratio of ram pressure to magnetic pressure:

\begin{align}
    \Lambda=\frac{2\dot{M} v}{B^2 r^2}
\end{align}
for both the stellar wind and the planetary outflow, where $\dot{M}$, $v$, and $B$ are defined at radial distance $r$.  For the stellar wind, we assume the speed is the same as it is for the Sun (400 km/s).  The speed does drop as one approaches the Sun, but not dramatically; \cite{venzmer_2018} predicted that the Parker Solar Probe would see a speed of 340 km/s at 0.16 AU and 290 km/s at 0.046 AU.  What Parker actually measured near perihelion on January 17, 2021, when it was 0.10 AU from the Sun, was 250--320 km/s.  To calculate the mass loss rate, we use the astrospheric observations of \cite{wood_2005b}.  This study found that the mass-loss rate scales with the X-ray flux as $\dot{M} \propto F_X^{1.34 \pm 0.18}$ until $F_X = 7 \times 10^{5}$ erg cm$^{-2}$ s$^{-1}$, at which point it abruptly falls from $\sim$80$\times$ solar to $\sim$8$\times$ solar.  This may be due to a large-scale change in magnetic topology, as extremely active stars tend to have a polar starspot while less active stars have more starspots at low latitudes.  HD 63433 has an X-ray flux of $F_X = 1.4 \times 10^{6}$ (cgs), which is just past the transition point between these two regimes.  We therefore adopt a value of 1.2\e{13} g/s, corresponding to the lower value of 8 times solar mass loss rate.  This value is similar to the 16$\times$ predicted by simulations in \cite{cranmer_2017}, but it is possible that the star's actual mass loss rate is closer to 80x solar.

Using this stellar mass loss rate, we can obtain an initial estimate for the ratio of ram to magnetic pressure if we assume that the stellar magnetic field is a dipole unaffected by the wind.  We find that $\Lambda_* = 43$ for b and $\Lambda_* = 690$ for c.  However, the very high $\Lambda_*$ means that the stellar magnetic field is carried with the wind and should be nearly radial at the positions of the two planets.  The stellar magnetic field at the location of the planet will therefore be much higher, and the corresponding $\Lambda_*$ will be much lower, than this simple dipole model predicts.

We next calculate the corresponding $\Lambda$ for the planets.   Planet c loses mass at a rate of 3\e{10} g/s in our fiducial model.  At 2 $R_p$, the outflow has a low ionization fraction and has not passed the sonic point.  At this point, $v \sim$ 2 km/s, giving $\Lambda_c = 10$.  Planet b loses mass at a rate of 8\e{10} g/s in our fiducial model.  At 2 $R_p$, the modeled outflow still has a low ionization fraction and has not passed the sonic point.  The outflow has a velocity of $v \sim$ 0.8 km/s, which translates to $\Lambda_b = 5$.  Our choice of scaling law for the planetary magnetic field means that $\Lambda \propto r^4$ at constant $v$, but since $v$ accelerates, $\Lambda$ rises even more steeply with $r$.   For comparison, a 0.4\% transit depth corresponds to a photospheric radius of 2.9 $R_p$ for b and 2.3 $R_p$ for c, while a 10\% transit depth corresponds to 15 and 12 $R_p$.  This means the outflow becomes ram pressure dominated inside of the triplet helium photospheric radius, and well inside of the sonic radius.  We conclude that the planetary magnetic field is probably insignificant in shaping the outflow for the fiducial magnetic field strengths.  However, if the magnetic fields are several times larger than fiducial--which is easily possible, given the poor theoretical knowledge of planetary dynamos--these $\Lambda$ values would decrease by a factor of 25-100, making the magnetic field highly significant in shaping the outflow.  Most simulations show that magnetic fields should decrease the outflow rate and the signal strength \citep{owen_2014,khodachenko_2015,arakcheev_2017}, although \cite{carolan_2021} predicts an increase in the outflow rate.

We can use this same approach to estimate the radial distance from the planet where the stellar wind begins to overwhelm the planetary outflow.  To compute this, we take the ratio of the ram pressures:

\begin{align}
    \Pi_{WW} = \frac{\dot{M_*}}{\dot{M_p}} \frac{v_*}{v_p} \frac{r^2}{a^2}
\end{align}
At 4 $R_p$, we obtain $\Pi_{WW} = 1.3$ for c and 3.2 for b.  These numbers are 3.0 and 1.6 at 15 $R_p$, indicating that the stellar wind plays a significant role in shaping the observed Ly$\alpha$ outflow.  This is fully consistent with our simulations (Subsection \ref{subsec:model_results}).

Lastly, we compute the effect of the interplanetary magnetic field on the outflow.  In the Parker model, the field is dragged along by the stellar wind, giving rise to a radial component that scales as $r^{-2}$ and an azimuthal component that scales as $r^{-1}$, pointing opposite to the direction of stellar rotation \citep{parker_1958}.  We use the measured radius dependence of the interplanetary field in the solar system \citep{hanneson_2020} to estimate the value of the stellar field at the positions of HD 63433b/c, assuming that the interplanetary field has the same proportionality with the stellar surface field in both systems.  For b, we find a field strength of 0.02 G and an angle $\theta = \arctan(\omega_* a / v_*) = 16^\circ$. 
For c, we find a field strength of 0.005 G and an angle of $31^\circ$.  HD 63433 has a magnetic field 5 times stronger than the Sun's, and the predicted field strength at the location of planet c is approximately a factor of five larger than the magnetic field strength recorded by the Parker Solar Probe's FIELDS instrument during its June 2020 perihelion, when it was 0.13 AU from the Sun (similar to c's 0.14 AU semimajor axis).  We therefore conclude that the distance scaling is in reasonable agreement with solar system observations.  However, the angles predicted by the Parker model are somewhat less reliable; in this model the solar interplanetary field at Earth's distance should have a value of $\theta=-40^\circ$, yet it is rarely within 10 degrees of that value.  There are also frequent deviations below -60$^\circ$ and above 50$^\circ$ at both solar maximum and minimum \citep{tasnim_2016}.  If we use these nominal values to compute the ratio of the planetary ram pressure to the stellar magnetic pressure, we find values of 42 for c and 4.3 for b at 2 $R_p$.  At 15 $R_p$, the ratios are 4.4 for c and 2.2 for b.  We conclude that the interplanetary magnetic field likely has some influence on the very outermost portions of the outflow, but it is probably less significant than the ram pressure from the stellar wind.


It is important to remember that these calculations are order-of-magnitude estimates and nearly all of the quantities involved have significant uncertainties.  The Sun's magnetic field is complex and variable; at Earth's distance, it has been shown to vary by a factor of a few over the course of the solar cycle.  The $B^2$ dependence of magnetic pressure means that a factor of a few uncertainty in the magnetic field strength translates to an order of magnitude uncertainty in the magnetic pressure.  Our estimated stellar wind mass-loss rate is based on a small number of indirect measurements in \cite{wood_2005b}.  It also falls close to a breakpoint in the scaling relation; it is possible that the mass-loss rate is closer to 80$\times$ solar than to 8$\times$ solar.  The wind velocity of stars other than the Sun is unknown.  There are no direct measurements of magnetic field strengths for extrasolar planets, and the field strengths of super-Earth cores may not scale simply as $\Omega^{1/2}$.  We utilize model predictions for the mass loss rates of the two planets, which may vary by a factor of a few depending on our starting assumptions.  Nevertheless, our fiducial values indicate that the stellar magnetic field is expected to be predominantly radial at the positions of both planets, with a significant tangential component pointing opposite the orbital direction.  They also suggest that the stellar wind will have a significant effect on the shape of the outflow, in good agreement with our 3D models.  Lastly, they indicate that the interplanetary magnetic field may influence the the outflow at the relatively large separations probed by our Ly$\alpha$ observations, and that the planetary magnetic field likely has a negligible effect.

\subsubsection{Radiation pressure}
Our outflow models also neglect to consider the effect of radiation pressure.  This effect has been invoked to explain detections of highly blueshifted Ly$\alpha$ absorption for multiple planets (e.g. \citealt{vidal-madjar_2003,ehrenreich_2012,bourrier_2014}).  However, we do not expect radiation pressure to be important in the HD 63433 system.  Since the radiation pressure comes predominantly from Ly$\alpha$, which is strongly absorbed by neutral hydrogen, we can calculate the radiation pressure using the Ly$\alpha$ flux inferred from observations.  Assuming that all species are coupled, we find that $P_{\rm rad} = F_{Ly\alpha}/c = 9 \times 10^{-8}$ dynes/cm$^2$ at the semimajor axis of planet c.  We compare this to the stellar wind ram pressure, $P_{\rm wind} = \dot{M}v/(8\pi a^2) = 4 \times 10^{-6}$ dynes/cm$^2$.  The ratio between the two is $P_{\rm wind}/P_{\rm rad} = 45$.  Since both pressures scale as $a^{-2}$, this ratio is the same for planet b.  The stellar wind ram pressure is the lowest pressure that exists in our simulation.  The total pressure is higher in the interaction region between the wind and the planetary outflow, and continues to increase as we move closer to the planet.  As a result, radiation pressure is even less important in altering the hydrodynamics.  

One might wonder if the ion-neutral collision cross section is high enough for the hydrodynamic assumption to be valid.  For collisions between H and H+, $\langle \sigma v \rangle = 3 \times 10^{-9}$ cm$^3$ s$^{-1}$ (\citealt{draine_2011}, their Table 2.1).  Assuming a v that corresponds to $10^4$ K and a characteristic length scale of $R_p$, the hydrodynamic assumption corresponds to $\rho >\sim 10^{-18}$ g cm$^{-3}$.  This density condition holds true for almost the entirely of the planetary outflow from both planets, although it begins to break down for c in the region outside the stream lines and the low-density interior of the extended tail.  The proton-proton collision cross section is substantially larger, and imposes a less limiting condition of $\rho >\sim 10^{-20}$ g cm$^{-3}$.  Our conclusions--that the hydrodynamic condition is mostly satisfied, and that radiation pressure is unlikely to significantly affect the photoevaporative winds--agree with those of \cite{debrecht_2020}, who used a 3D hydrodynamic simulation to study the outflow from the hot Jupiter HD 209458b.  However, we note that the hydrodynamic condition is not satisfied in the stellar wind in either our simulation or \cite{debrecht_2020}, and that more sophisticated modelling would be useful for the interaction region between the stellar wind and the planetary outflow.


\section{Conclusion}
\label{sec:conclusion}
In this study we use HST and Keck to look for escaping hydrogen and helium from two mini Neptunes orbiting a young solar analogue.  We detect $11.5 \pm 1.5$ \% Ly$\alpha$ absorption in the blue wing during a transit of planet c, but not during a transit of planet b.  We do not detect excess helium absorption during transits of either planet, to a stellar-variability-limited upper limit of $\sim$0.5\%.  We use near-contemporaneous XMM-Newton data to characterize the stellar high-energy environment during our Ly$\alpha$ and helium observations, and combine these observations with ground-based optical monitoring data and archival ROSAT data to constrain the long-term stellar activity cycle.  We use the measured stellar X-ray spectrum and an extrapolated extreme UV spectrum as inputs to 3D hydrodynamic models of hydrogen- and helium-rich planetary outflows, which we compare to the observational data.

For c, our hydrogen-rich models provide a reasonable match to the measured shape and depth of the transit light curve in the blue wing of the Ly$\alpha$ line.  Similarly, their predictions for the magnitude of absorption in the metastable helium line are consistent with our upper limit for this planet.  However, the observed blue wing absorption is less compact in velocity space than the simulated absorption.  We speculate that this might be due to charge exchange with the stellar wind, which is difficult to model accurately with our hydrodynamic model, but which \cite{tremblin_2013} concludes can cause 10\% Ly alpha absorption at 100 km/s.  In addition, our models do not account for the effects of magnetic fields.  While the effect of the planetary magnetic field on the outflow is likely negligible, the interplanetary field could play a subdominant role in confining and guiding the outflow beyond $\sim10 R_p$.

Our observations and models both suggest that planet b is fundamentally different from planet c.  Our models predict that if planet b hosts a hydrogen-rich atmosphere it should also exhibit strong Ly$\alpha$ absorption during transit, but this is definitively ruled out by our observations.  Our models also predict a detectable metastable helium absorption signal from planet b, which is inconsistent with our observational upper limit.  The predicted mass loss timescale for planet c is longer than the age of the system, but the corresponding mass loss timescale for planet b is significantly shorter.  This implies that c could have retained a primordial H/He atmosphere, while b probably does not.  If b is a rocky core, it would have to be unusually massive, but a water-rich composition with a high mean molecular weight atmosphere could explain both the radius and the Ly$\alpha$ non-detection.  Fortunately, HD 63433 is a bright nearby star, and its planets are favorable targets for atmospheric characterization by HST and JWST.  If planet b does host a hydrogen-dominated atmosphere, it may have detectable absorption from water and other molecules in its infrared transmission spectrum.  If b hosts a high molecular weight atmosphere, detecting it in outflow may still be possible by looking at metal FUV lines \citep{garcia_2021}.

\textit{Software:}  \texttt{numpy \citep{van_der_walt_2011}, scipy \citep{virtanen_2020}, matplotlib \citep{hunter_2007}, stan\citep{stan2018}, stistools, SAS, HEAsoft}

\acknowledgments
This study was based on observations with the NASA/ESA Hubble Space Telescope, obtained from the data archive at the Space Telescope Science Institute. STScI is operated by the Association of Universities for Research in Astronomy, Inc., under NASA contract NAS5-26555. Support for this work was provided by NASA through grant number GO-16319 from STScI.  This study also utilized data obtained at the W. M. Keck Observatory, which is operated as a scientific partnership among the California Institute of Technology, the University of California and the National Aeronautics and Space Administration. The Observatory was made possible by the generous financial support of the W. M. Keck Foundation.   L. dos Santos and D. Ehrenreich acknowledge that this project received funding from the European Research Council (ERC) under the European Union's Horizon 2020 research and innovation programme (project {\sc Four Aces} grant agreement No 724427), and it has been carried out in the frame of the National Centre for Competence in Research PlanetS supported by the Swiss National Science Foundation (SNSF). TGW acknowledges support from STFC consolidated grant number ST/R000824/1.  SH acknowledges CNES funding through the grant 837319.  GWH acknowledges long-term support of the APT program from NASA, NSF, Tennessee State University, and the State of Tennessee through its Centers of Excellence Program.

\appendix
In the published paper, we will make publicly available the reconstructed stellar spectrum, the reduced observations in both Ly$\alpha$ and helium, and the 1D profiles from our hydrodynamic model.

\section{Alternate analyses of Ly$\alpha$ data}
\label{section:appendix_alt_analyses}
The fiducial analysis reported in the previous subsection was performed by the first author (Michael Zhang, MZ).  Two other analyses were performed by two co-authors (Luca Fossati, LF and Leonardo dos Santos, LDS) using independent pipelines, and with no communication other than agreeing on a common velocity range in which to look for absorption: [$-140$,$-10$] km/s in the blue wing, [100,200] km/s in the red. 

In LDS's analysis, he leveraged the time tag stream of events in the raw datasets to break down each HST orbit into ten subexposures to improve the temporal resolution of the time series. The subexposures are then reduced using \texttt{calstis} with its default settings.  In order to correct for telescope breathing, he calculated the total flux between 1200-1248~\AA\ (excluding the geocoronal contamination range) for each subexposure, and subsequently phase-folded the flux time-series to the orbital period of the space telescope ($P_\mathrm{\it HST} = 95.42$~min). He tried to correct for the breathing effect by fitting this time series to a Fourier decomposition model for the systematic modulation with varying degrees and period equal to that of HST. This is the same approach used by \citet{bourrier_2017b} and references therein. For both visits of planet b, and the first visit of c, a first degree systematics model is favored over higher degree models and no systematics models by a $\Delta$BIC $>$ 10. A model with no systematics is favored for the second visit of c, since higher degrees do not significantly improve the BIC statistic.  The uncertainties are calculated assuming that the observations are in the Poisson counting regime (see \citet{dos_santos_2021}). LDS did not correct for potential modulation due to stellar activity, which is expected for young stars, in this analysis.  This is very difficult to do without simultaneous observations at other wavelengths.

In LF's analysis, he first considered the calibrated 2-dimensional spectra providing information on the photon arrival time (``\_tag.fits'' files) and split each HST observation into 5 sub-exposures of equal exposure time. From each sub-exposure image, he extracted the stellar spectra using a slanted rectangular extraction box with an aperture of 40 pixels and the background employing an identical extraction box, but shifted upwards by 130 pixels. He then removed the relative background from the stellar spectra, phased each sub-exposure with HST's orbit, and tried to correct for the breathing effect.  To do this, he excluded the observations obtained during the first HST orbit of each visit, because the first HST orbit is notoriously affected by additional systematics. He modelled the breathing effect separately for each HST visit as a polynomial of varying order, selecting the one that minimizes the Bayesian Information Criterion: BIC = $\chi^2 + k \log{N}$, where $k$ is the number of free parameters and $N$ is the number of data points. For the first and second visit covering the transit of planet b, he described the breathing effect with a second and third order polynomial, respectively, while for the first and second visit covering the transit of planet c, he described the breathing effect with a second and first order polynomial, respectively. For each HST visit, which was characterised by a distinct shape for the breathing effect, he applied the same correction to the Ly$\alpha$ fluxes for the blue and red wings. 

All three authors found that breathing corrections do not significantly or consistently reduce the scatter of the orbit-averaged fluxes.  MZ considered these corrections in his preliminary analyses but did not feel that they were justified.  LF found that his corrections have virtually no effect on the blue wings, but reduce the scatter in the red wing of planet b and increase the scatter in the red wing of planet c.  LDS finds that his corrections marginally increase the scatter for both wings of planet c, but marginally decrease the scatter for both wings of planet b.  Given these results, we decided to exclude breathing corrections from the fiducial analysis presented in this study.

\begin{figure*}
  \centering 
  \subfigure {\includegraphics
    [width=0.5\textwidth]{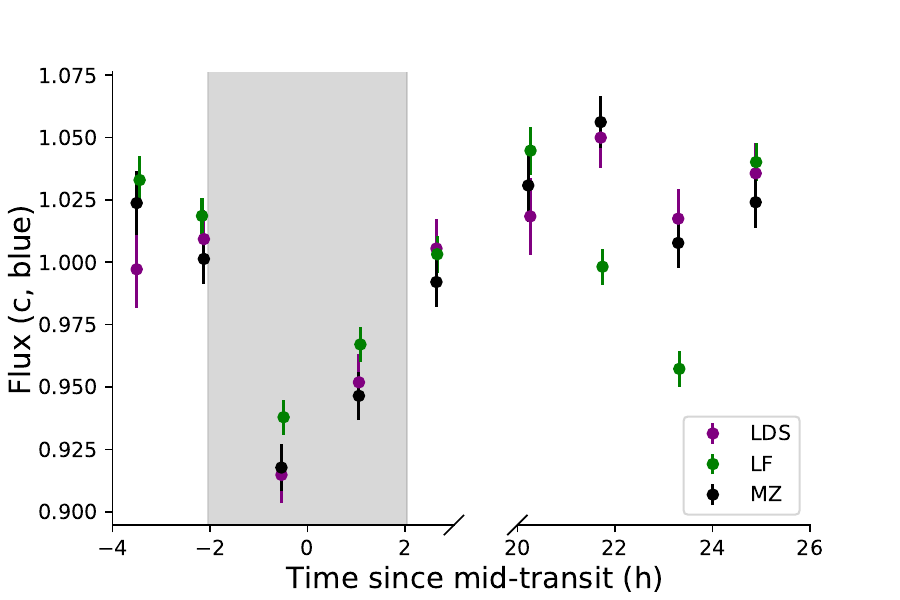}}\subfigure {\includegraphics
    [width=0.5\textwidth]{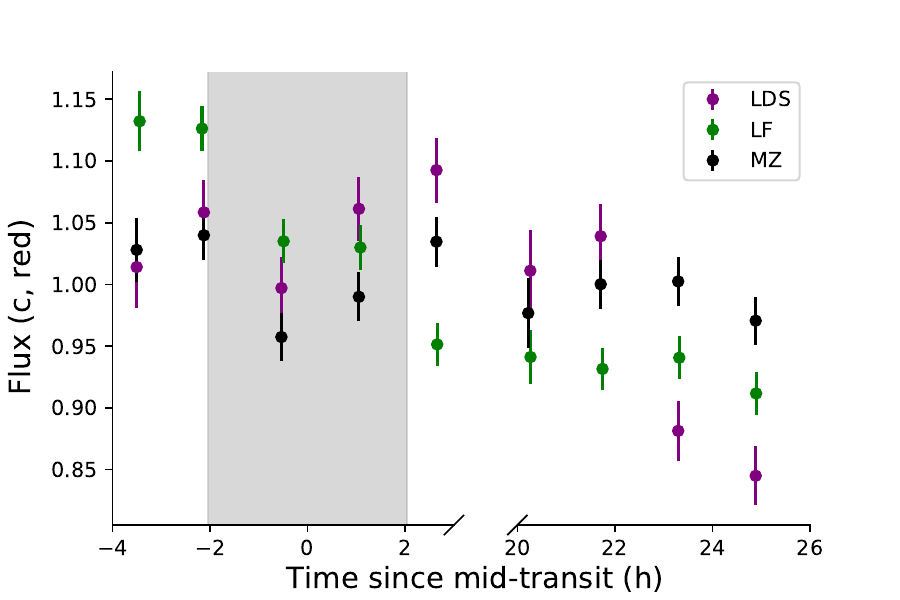}}
  \subfigure {\includegraphics
    [width=0.5\textwidth]{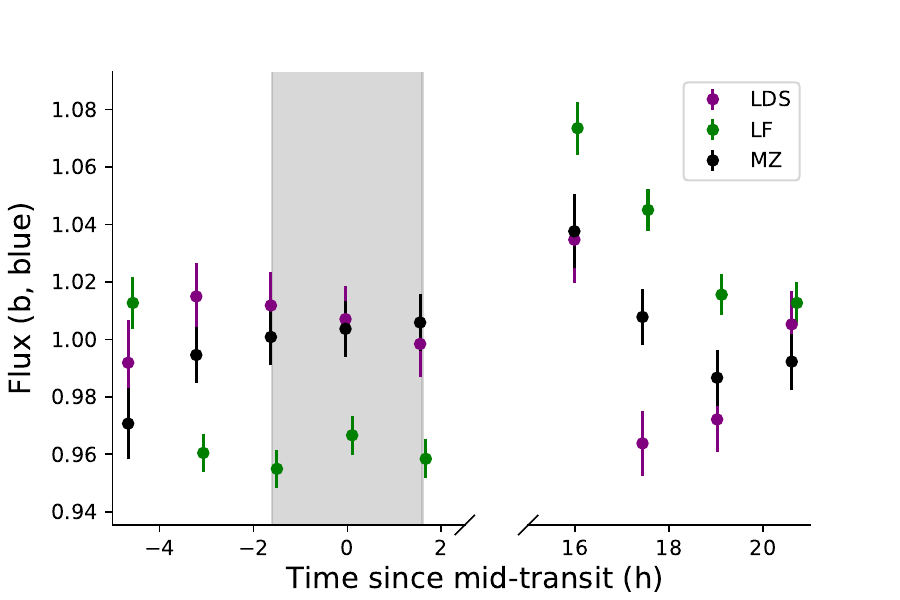}}\subfigure {\includegraphics
    [width=0.5\textwidth]{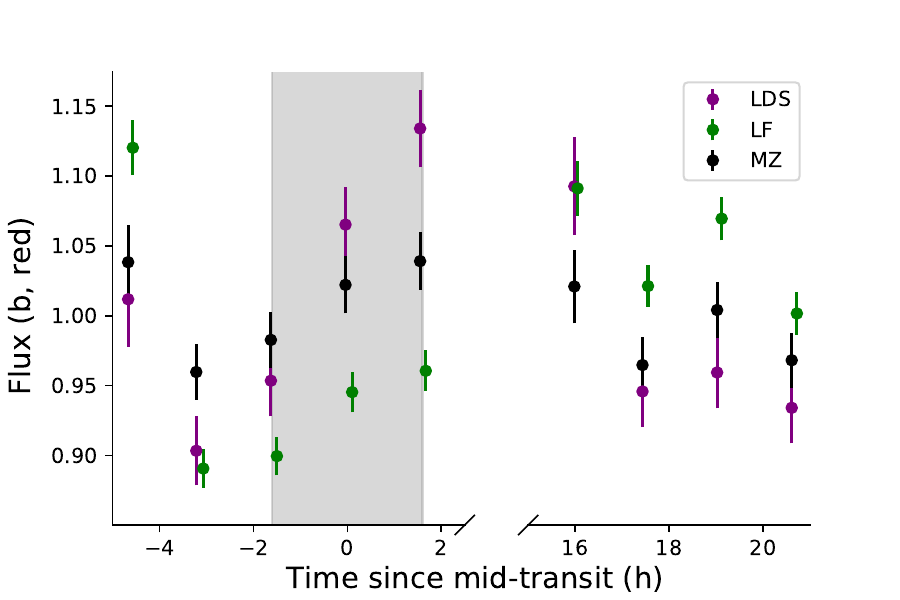}}
    \caption{Comparison of independent analyses by MZ, LDS, and LF.  The blue wing light curves are shown on the left, and the red wing light curves are shown on the right.  Planet c is shown in the top panel, and planet b is shown in the bottom panel.  The duration of the white-light transit is indicated by the grey shaded region.  We exclude the data from the failed second observation of planet b in this comparison.}
\label{fig:comp_analyses}
\end{figure*}

\begin{figure*}
  \centering 
  \subfigure {\includegraphics
    [width=0.5\textwidth]{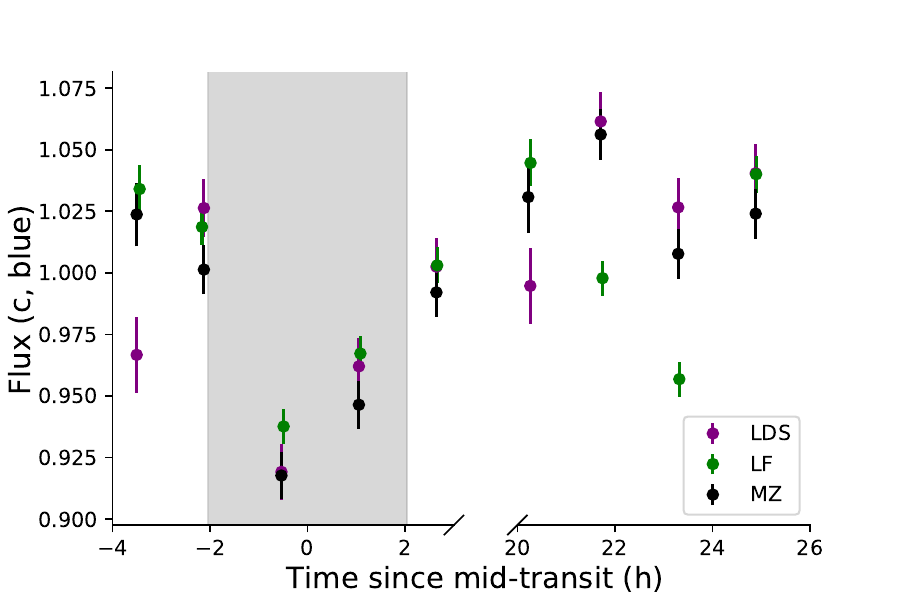}}\subfigure {\includegraphics
    [width=0.5\textwidth]{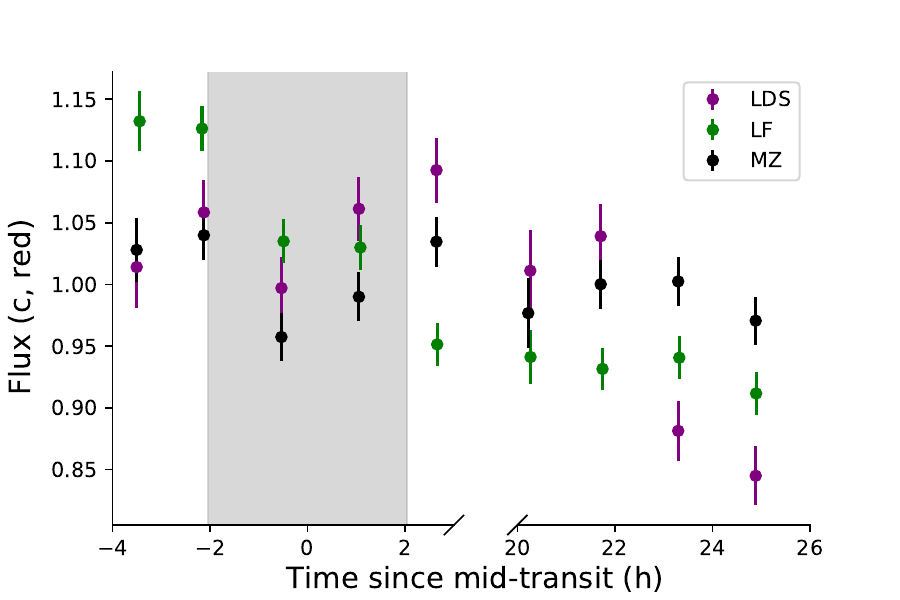}}
  \subfigure {\includegraphics
    [width=0.5\textwidth]{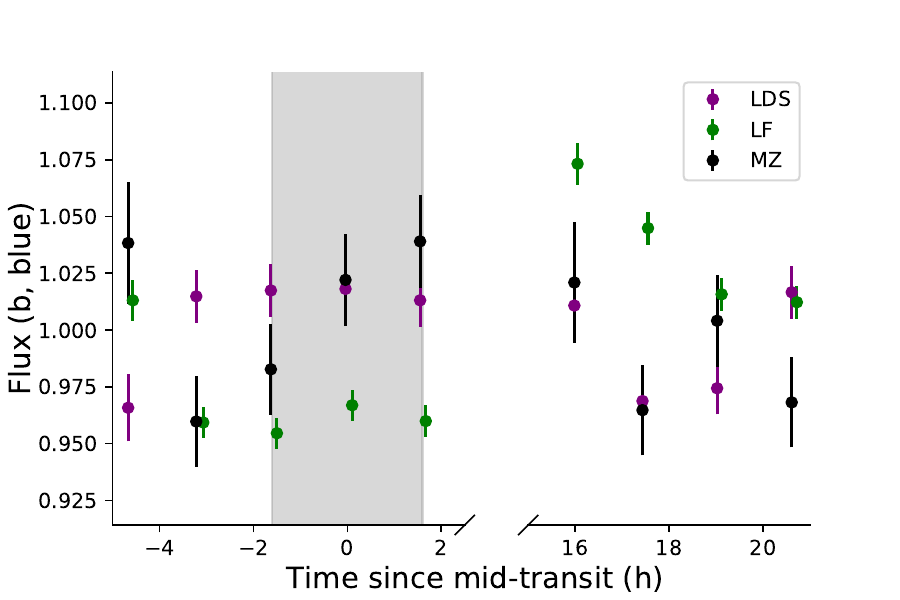}}\subfigure {\includegraphics
    [width=0.5\textwidth]{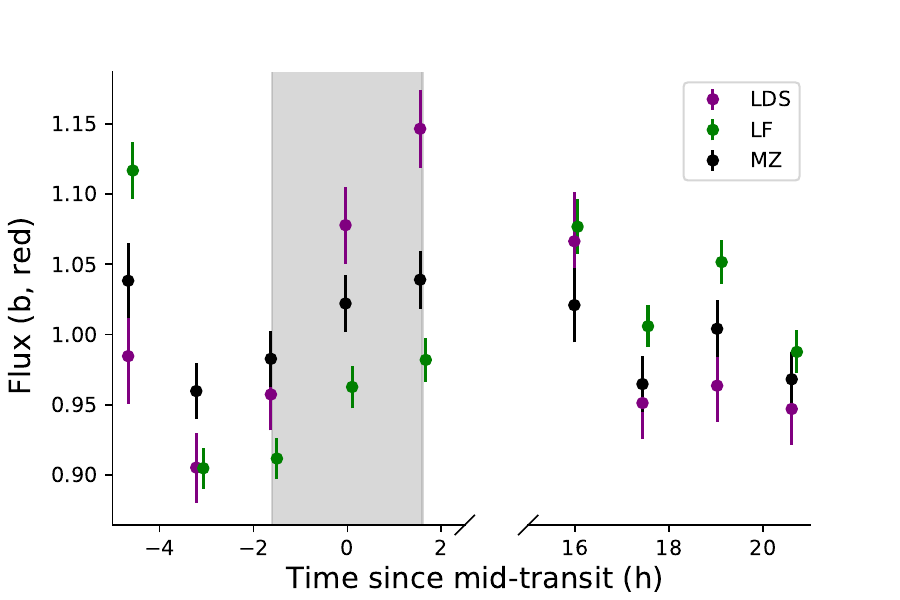}}
    \caption{Same as Figure \ref{fig:comp_analyses}, but breathing corrections are included in the LF and LDS analyses.  The fiducial analysis remains the same.}
\label{fig:breathing_comp_analyses}
\end{figure*}

Figure \ref{fig:comp_analyses} compares the light curves from the three analyses, all of which are shown with no breathing correction.  The results are largely consistent, although we find that the fiducial analysis has a lower average scatter than the others.  All three analyses show clear blue wing absorption during the transit of planet c.  All three analyses find no evidence for absorption in either wing during the transit of b.  The fiducial analysis shows tentative red wing absorption from c while the other two do not.  This difference may be due to the significantly higher standard deviation in the light curves from the two alternative analyses (2.8\% vs. 7.9\% for LDS and 8.0\% for LF).  The conclusions above do not change when the fiducial analysis is compared to the breathing-corrected versions of LF's and LDS' analyses (Figure \ref{fig:breathing_comp_analyses}).

\bibliographystyle{apj} \bibliography{main}

\end{document}